\documentclass{JHEP3}

\usepackage{amsmath,amssymb}           
\usepackage{epsfig}                    
\usepackage[matrix,arrow]{xy}          
\usepackage{xspace,bbold}              
\usepackage[vcentermath]{youngtab}     

\allowdisplaybreaks                    

\newcommand{\bq}{\begin{equation}}
\newcommand{\eq}{\end{equation}}
\newcommand{\bea}{\begin{eqnarray}}
\newcommand{\eea}{\end{eqnarray}}

\newcommand{\dd}{\mathrm{d}}
\newcommand{\ee}{\mathrm{e}}
\newcommand{\ii}{\mathrm{i}}

\newcommand{\bbZ}{\mathbb{Z}}
\newcommand{\bbR}{\mathbb{R}}
\newcommand{\bbC}{\mathbb{C}}

\DeclareMathOperator{\SU}{\mathit{SU}}
\DeclareMathOperator{\SO}{\mathit{SO}}
\DeclareMathOperator{\SL}{\mathit{SL}}
\DeclareMathOperator{\GL}{\mathit{GL}}
\DeclareMathOperator{\Symp}{\mathit{Sp}}
\DeclareMathOperator{\Spin}{\mathit{Spin}}

\DeclareMathOperator{\symp}{\mathit{sp}}

\DeclareMathOperator{\Cliff}{Cliff}

\newcommand{\rep}[1]{\mathbf{#1}}
\newcommand{\id}{\mathbb{1}}

\DeclareMathOperator{\tr}{tr}

\DeclareMathOperator{\re}{Re}

\DeclareMathOperator{\E7}{\mathit{E}_{7(7)}}
\newcommand{\hg}{{\SU(8)/\bbZ_2}}

\newcommand{\met}{\eta}
\newcommand{\ksp}{\zeta}

\DeclareMathOperator{\Diff}{Diff}

\newcommand{\smu}{{*\mu}}

\newcommand{\tgam}{\hat{\gamma}}
\newcommand{\tg}{\hat{g}}

\newcommand{\tA}{\tilde{A}}

\newcommand{\FF}{\mathcal{F}}
\newcommand{\tFF}{\tilde{\FF}}
\newcommand{\tL}{\tilde{\Lambda}}
\newcommand{\tphi}{\tilde{\varphi}}

\newcommand{\SG}{G_{\text{closed}}(M)}

\newcommand{\Lab}{\Lambda_{(\alpha\beta)}}
\newcommand{\tLab}{\tL_{(\alpha\beta)}}

\newcommand{\mukai}[2]{\big<{#1},{#2}\big>}

\newcommand{\Leo}{\Lambda^\text{even/odd}}

\newcommand{\stt}{{\SU(3)\times\SU(3)}}

\newcommand{\gfive}{\ii\gamma_{(4)}}

\title{M-theory, exceptional generalised geometry and superpotentials}

\author{Paulo Pires Pacheco \\
   Department of Physics,
   Imperial College London \\
   London, SW7 2AZ, UK}
\author{Daniel Waldram \\
   Department of Physics,
   Imperial College London \\
   London, SW7 2AZ, UK \smallskip \\
   Institute for Mathematical Sciences,
   Imperial College London \\
   London, SW7 2AZ, UK}

\abstract{
We discuss the structure of ``exceptional generalised geometry''
(EGG), an extension of Hitchin's generalised geometry that provides a
unified geometrical description of backgrounds in eleven-dimensional
supergravity. On a $d$-dimensional background, as first described by
Hull, the action of the generalised geometrical $O(d,d)$ symmetry
group is replaced in EGG by the exceptional U-duality group
$E_{d(d)}$. The metric and form-field degrees of freedom
combine into a single geometrical object, so that EGG naturally
describes generic backgrounds with flux, and there is an EGG analogue
of the Courant bracket which encodes the differential geometry. Our
focus is on the case of seven-dimensional backgrounds with $N=1$
four-dimensional supersymmetry. The corresponding EGG is the
generalisation of a $G_2$-structure manifold. We show it is
characterised by an element $\phi$ in a particular orbit of the
$\rep{912}$ representation of $\E7$, which defines an
$\SU(7)\subset\E7$ structure. As an application, we derive the generic
form of the four-dimensional effective superpotential, and show that
it can be written in a universal form, as a homogeneous
$\E7$-invariant functional of $\phi$.
}

\preprint{Imperial/TP/08/DW/01}

\begin{document}


\section{Introduction}
\label{sec:intro}


Type II string backgrounds in $d$ dimensions which include non-trivial
fluxes have a natural description in terms of Hitchin's generalised
geometry~\cite{GCY,Gualtieri,H-GK,H-brack}, where the metric and
NS--NS $B$-field are combined into a single geometrical object,
transforming under $O(d,d)$. This description has proved very useful
in, among other things: characterising supersymmetric backgrounds,
finding new examples with non-zero fluxes and writing supersymmetric
low-energy effective theories~\cite{GLMW}--\cite{BFMMPZ}; describing
topological string theories and generic $N=(2,2)$
$\sigma$-models~\cite{Kap}--\cite{LSSW}; as well as motivating the
structure of non-geometrical backgrounds~\cite{Tfold}--\cite{Hull:2007jy}.

The aim of this paper is to understand some details of how similar
constructions based on the exceptional groups $E_{d(d)}$ can be used to
describe M-theory, or more precisely, eleven-dimensional supergravity
backgrounds. The general form of such constructions, as well as
those arising from type II theories, has been described recently by
Hull~\cite{chris}. That work was partially motivated by the existence
of so-called ``non-geometrical'' backgrounds, which appear consistent
as string theory vacua and are typically dual to supergravity
backgrounds, but do not themselves have a consistent global
description in supergravity. This suggested considering extensions of
generalised geometry based on generic $O(d,d)$ or $E_{d(d)}$
bundles, whereas only a subclass of such bundles arise in
supergravity.  The corresponding geometry was generically dubbed
``extended'' (or more specifically ``M-geometry'' for the
generalisation of eleven-dimensional supergravity).

Here, however, we will consider only supergravity backgrounds and
concentrate on the physically important example of seven-dimensional
backgrounds and hence the group $\E7$. There are two parts to the
analysis. First to build the analog of the generalised geometry
and then to describe the geometrical objects that characterise
supersymmetric $N=1$ backgrounds in four dimensions. In analogy to the
type II analysis in~\cite{GLW1,GLW2}, we consider the particular
application of writing the $N=1$ superpotential in a generic
$\E7$-invariant form.

The essential idea of the construction is that $O(d,d)$ symmetries of
generalised $d$-dimensional geometry, which in string theory are
related to T-duality symmetries, should be replaced by the U-duality
exceptional symmetry groups $E_{d(d)}$~\cite{CJ,Udual,HT}. Since the
U-duality connects all the bosonic degrees of freedom of
eleven-dimensional supergravity, or for type II theories, both NS--NS
and R--R degrees of freedom, this extension should provide a
geometrisation of generic flux backgrounds. The fact that the full
eleven-dimensional supergravity could be reformulated in terms of
$E_{d(d)}$ objects was first pointed out by de~Wit and
Nicolai~\cite{deWN} (for $\E7$) and then by Nicolai~\cite{Nic} and
Koepsell, Nicolai and Samtleben~\cite{KNS} (for $E_{8(8)}$), the
latter calling the construction ``exceptional geometry''. Motivated by
these authors' and Hull's nomenclature~\cite{chris}, we will refer
to the variant of Hitchin's generalised geometry relevant to
eleven-dimensional (and type II) supergravity as EGG for ``exceptional
generalised geometry''. This reserves ``extended'' and ``M'' for
generic M-theory backgrounds, potentially including non-geometrical
examples.

We should note here that recently there have also been more ambitious
related proposals connecting infinite-dimensional exceptional algebras
to supergravity. The original proposal of~\cite{West1,West2} had the
goal of giving an eleven-dimensional covariant formulation of M-theory
with $E_{11}$ invariance, as well as of lower-dimensional gauged
supergravities~\cite{RW}, while the work of~\cite{Nicolai} describes a
gauged-fixed version of the supergravity dynamics in terms of an
explicit $E_{10}$ coset construction.

The paper is arranged as follows. In section~\ref{sec:EGG}, after
briefly reviewing generalised geometry, we describe how the analogous
$\E7$-invariant EGG can be defined for a seven-dimensional manifold
$M$. Much of this analysis appears in~\cite{chris}. The
corresponding exceptional generalised tangent space (EGT) encodes all
the topological information of the conventional tangent space $TM$ as
well as the topology of the ``gerbe'', the analogue of a
$U(1)$-bundle, on which the supergravity three-form $A$ is a
connection. We give new details on how the tangent space bundle
embeds into the EGG, as well as the precise form of the gerbes.
We also define the analogue of the Courant bracket on the
EGT which encodes the differential geometry. We then introduce the
notion of an $\hg$ structure on the EGT and show that this encodes
supergravity metric $g$ and three-form $A$. (This is equivalent to the
result~\cite{chris}, familiar from toroidal reductions~\cite{CJ}, that
they parameterize an element of the coset space $\E7/(\hg)$.) We show
in particular how the non-linear Chern--Simons-like terms for $A$
appear naturally in this formalism.

In section~\ref{sec:SEGG}, we again start with review. We discuss the
pure $O(6,6)$ spinors $\Phi^\pm$ which characterise generalised
geometrical six-dimensional type II backgrounds preserving $N=2$
supersymmetry in four dimensions, and define an $\stt$ structure on
the generalised tangent space. We then make the analogous analysis for
EGG. (A short related discussion of supersymmetric backgrounds
appeared in section~8 of~\cite{chris}.) Without flux, for the
background to have $N=1$ supersymmetry in four dimensions, the
compactification manifold must have a $G_2$-structure. We show that
the generalisation to EGG is that the EGT must admit an $\SU(7)$
structure. This is equivalent to the existence of a nowhere vanishing
tensor $\phi$ transforming in the $\rep{912}$ of $\E7$. It is not a
generic element but, to be stabilized by $\SU(7)$ must lie in a
particular orbit under $\E7$.

In section~\ref{sec:W} we give an application of the these
results. We first derive the form of the superpotential for a generic
$N=1$ compactification of eleven-dimensional supergravity slightly
extending the results of~\cite{HM}. We then use the structure $\phi$,
which plays the role of the chiral scalar fields in the
four-dimensional theory, to show that the superpotential can be
rewritten in a $\E7$-invariant form.

We conclude in section~\ref{sec:concl} with a brief discussion of our
results and of possible further work.


\section{EGG in seven dimensions}
\label{sec:EGG}

In this section we discuss the structure of exceptional generalised
geometry, focusing on the case of seven dimensions. Much of this
analysis also appears in~\cite{chris}. Here we approach the construction
from the point of view of geometrical structures and include some new
details on the gerbe topology and tangent space embedding, as well the
definition of the analogue of the Courant bracket. Our
particular motivation will be compactification of eleven-dimensional
supergravity to a four-dimensional effective theory. However here we
will first simply introduce the EGG formalism and leave the more
detailed connection to supergravity to the following section. We start
by reviewing the structure of generalised
geometry~\cite{GCY,Gualtieri,H-GK,H-brack} since the EGG structure
arises in a very analogous way.


\subsection{Review of generalised geometry}
\label{sec:revGG}

In defining generalised geometry~\cite{GCY,Gualtieri,H-GK,H-brack} on
a $d$-dimensional manifold, one starts with the (untwisted)
generalised tangent space $E_0=TM\oplus TM^*$. Let us denote elements
of $E_0$ by $X=x+\xi$. There is then a natural $O(d,d)$-invariant
metric $\met$ on $E$ given by
\begin{equation}
   \met(X,X) := i_x\xi .
\end{equation}
In particular, $\met$ is invariant under the $\GL(d)\subset O(d,d)$
action on the fibres of $TM$ and $T^*M$. The metric is also invariant
under so called ``$B$-shifts'' where for any $B\in\Lambda^2T^*M$ we define
\begin{equation}
\label{eq:B-shift}
   \ee^B X := x + \left(\xi-i_xB\right) .
\end{equation}

The differential geometry of the generalised tangent space is encoded
in the Courant bracket which generalises the Lie bracket between two
vectors. It is defined by
\begin{equation}
\label{eq:Courant}
   [ x+\xi, y+\zeta] = [x,y] + \mathcal{L}_x\zeta - \mathcal{L}_y\xi
      - \tfrac{1}{2}\dd\left(i_x\zeta - i_y\xi\right) ,
\end{equation}
where $[x,y]$ is the usual Lie bracket of vector fields and
$\mathcal{L}_x$ is the Lie derivative, so
$\mathcal{L}_x\alpha=i_x\dd\alpha+\dd i_x\alpha$ for any form
$\alpha$. Note that its automorphism group is not the full group of local
$O(d,d)$ transformations but only the subgroup generated by
diffeomorphisms and the $B$-shifts~\eqref{eq:B-shift} (with $\dd
B=0$). Formally we can write this as the semidirect product
$\Diff(M)\ltimes\Omega^2_{\text{closed}}(M)$.

The usefulness of generalised geometry in string theory and
supergravity is that the ordinary metric $g$ and NS--NS two-form field
$B$ combine naturally into a single object, the so-called generalised
metric $G$~\cite{Gualtieri}. This is an $O(2d)$-invariant metric which
is compatible with $\met$, that is $\met^{-1}G\met^{-1}=G^{-1}$. If we
split into $TM$ and $T^*M$, we can write the $O(d,d)$ metric as the
matrix
\begin{equation}
   \met = \frac{1}{2}\begin{pmatrix}
         0 & \id \\ \id & 0
      \end{pmatrix} ,
      \qquad
      \text{where} \qquad
   X = \begin{pmatrix} x \\ \xi \end{pmatrix} .
\end{equation}
The generic generalised metric can then be written as
\begin{equation}
\label{eq:gen-G}
   G = \frac{1}{2}\begin{pmatrix}
         g - Bg^{-1}B & Bg^{-1} \\
         - g^{-1}B & g^{-1}
      \end{pmatrix} ,
\end{equation}
where $g$ is an ordinary Riemannian metric, and
$B\in\Lambda^2T^*M$ is the NS-NS two-form.

Note that one can also define $G$ as a product structure $\Pi$. Writing
$\Pi=\met^{-1}G$ one has $\Pi^2=\id$, and in addition
$\met(X,Y)=\met(\Pi X,\Pi Y)$. Hence $\Pi$ is a product structure,
compatible with $\met$, and the projections $\frac{1}{2}(\id\pm\Pi)$
project onto two $d$-dimensional subspaces $C^\pm$, such that
$E=C^+\oplus C^-$. From this perspective, given $\met$, one sees that
$G$ defines an $O(d)\times O(d)$ structure, and $\met$ and $G$
decompose into the separate metrics on $C^+$ and $C^-$. One can view
$g$ and $B$ as parametrising the coset space $O(d,d)/O(d)\times
O(d)$.

One can also write
\begin{equation}
\label{eq:B-shiftG}
   G(X,Y) = G_0(\ee^B X, \ee^B Y)
\end{equation}
where
\begin{equation}
   G_0 = \frac{1}{2} \begin{pmatrix}
             g & 0 \\ 0 & g^{-1}
          \end{pmatrix} .
\end{equation}
If $B$ leads to a non-trivial flux $H$ it is only locally defined as a
two-form. Globally one must patch by gauge transformations, so on the
overlap $U_{(\alpha)}\cap U_{(\beta)}$ one has
\begin{equation}
   B_{(\alpha)} - B_{(\beta)} = \dd\Lambda_{(\alpha\beta)} ,
\end{equation}
(so $\Lambda_{(\alpha\beta)}=-\Lambda_{(\beta\alpha)}$) while on the
triple overlap $U_{(\alpha)}\cap U_{(\beta)}\cap U_{(\gamma)}$
\begin{equation}
   \Lambda_{(\alpha\beta)} + \Lambda_{(\beta\gamma)}
         + \Lambda_{(\gamma\alpha)}
      = \dd\Lambda_{(\alpha\beta\gamma)} ,
\end{equation}
(implying
$\Lambda_{(\alpha\beta\gamma)}=-\Lambda_{(\alpha\gamma\beta)}$
etc.). Mathematically this means $B$ is a connection on a gerbe (see
for instance~\cite{gerbes}). If the flux is quantised $H\in
H^3(M,\bbZ)$ then one has
$g_{(\alpha\beta\gamma)}=\ee^{\ii\Lambda_{(\alpha\beta\gamma)}}\in
U(1)$ and these elements satisfy a cocycle condition on
$U_{(\alpha)}\cap U_{(\beta)}\cap U_{(\gamma)}\cap U_{(\delta)}$
\begin{equation}
   g_{(\beta\gamma\delta)}g^{-1}_{(\alpha\gamma\delta)}
      g_{(\alpha\beta\delta)}g^{-1}_{(\alpha\beta\gamma)} = 1 .
\end{equation}
Formally the $g_{(\alpha\beta\gamma)}$ define the gerbe, while
the $\Lambda_{(\alpha\beta)}$ define a ``connective structure'' on the
gerbe. Together they encode the analogue of the topological data of a
$U(1)$ gauge bundle.

If $B$ is non-trivial, the form of the generalised
metric~\eqref{eq:B-shiftG} means that $G$ cannot really be an inner
product on sections of $E_0=TM\oplus T^*M$. Instead, the generalised
vectors must be sections of an extension $E$
\begin{equation}
\label{eq:gen-T}
   0 \longrightarrow T^*M \longrightarrow E
      \longrightarrow TM \longrightarrow 0 ,
\end{equation}
where on the intersection of two patches $U_{\alpha}\cap U_{\beta}$ one
identifies $X_{(\alpha)}=\ee^{-\dd\Lab}X_{(\beta)}$, or in components
\begin{equation}
\label{eq:patch}
   x_{(\alpha)} + \xi_{(\alpha)}
     = x_{(\beta)} + ( \xi_{(\beta)}
       + i_{x_{(\beta)}}\dd\Lambda_{(\alpha\beta)} ) .
\end{equation}
Here $\Lambda_{(\alpha\beta)}$ is the same one-form that appears in
defining the connection $B$ (though of course it is independent of the
particular choice of $B$). Thus we see that $E$ encodes both the
topological structure of the tangent space $TM$ and the connective
structure of the gerbe. Note that the form of the
twisting~\eqref{eq:patch} is such that, since the Courant bracket is
$B$-shift invariant (when $B$ is closed), it can still be defined on
sections of the twisted $E$.


\subsection{The exceptional generalised tangent space and $\E7$}
\label{sec:EGT}

We would now like to describe an exceptional generalised geometry (EGG),
analogous to the generalised geometry of Hitchin, but relevant to the
description of eleven-dimensional supergravity rather than simply the
NS--NS sector of type II. We will concentrate on the case of a
seven-dimensional manifold $M$. The basic construction has been
described, in general dimension $d$, in~\cite{chris} and is closely
related to the work of~\cite{deWN,Nic,KNS}.

Introducing the generalised tangent space allowed one to construct
objects transforming under $O(d,d)$. The $g$ and $B$ degrees of
freedom, parametrising the generalised metric $G$, then define an
element in the $O(d,d)/O(d)\times O(d)$ coset. This coset structure is
familiar from the moduli space of toroidal compactifications of NS-NS
sector ten-dimensional supergravity to $10-d$ dimensions~(see for
instance~\cite{Tduality}). In addition, there is, of course, a stringy
$O(d,d;\bbZ)$ T-duality symmetry relating equivalent
compactifications. The string winding and momentum charges transform
in the $2d$-dimensional vector representation of $O(d,d)$, namely
$TM\oplus T^*M$.

If one includes Ramond--Ramond fields, or equivalently considers
toroidal compactifications of eleven-dimensional
supergravity~\cite{CJ,Udual}, the moduli spaces are cosets
$E_{d(d)}/H$ (where $E_{d(d)}$ is the maximally non-compact real form
of the exceptional group and $H$ is the corresponding maximal compact
subgroup). Elements in the coset are parametrised by the components of
the eleven-dimensional supergravity fields on the compact space,
namely the metric $g$ and three-form $A$ potential (and potentially its dual
six-form $\tA$). The discrete subgroup $E_{(d(d)}(\bbZ)$ is the
U-duality group relating different equivalent M-theory
backgrounds~\cite{HT}. The momentum and brane charges fill out a
particular representation of $E_{d(d)}$. In particular, in $d=7$, the
momentum, membrane, fivebrane and Kaluza--Klein monopole
charges~\cite{KKmono} fill out the $\rep{56}$ representation of
$\E7$. (Note that the definition of $\E7$, together with some details of its
various representations, is summarised in appendix~\ref{app:E7}.)

Given this extension from the T-duality group $O(d,d)$ to U-duality
group $E_{d(d)}$, it is natural to introduce a corresponding extension
of generalised geometry. For $d=7$, the analogue of the generalised
tangent space is an ``exceptional generalised tangent space'' (EGT)
which transforms in the $\rep{56}$ representation of $\E7$. As in
generalised geometry, the $\GL(7,\bbR)$ structure group of the tangent
and cotangent spaces should be a subgroup of $\E7$, and we also expect
the gauge transformations of $A$, like the $B$-shifts in generalised
geometry, to be somehow embedded into $\E7$.

The construction of the EGT is as follows. As described in
appendix~\ref{app:E7def}, there is an $\SL(8,\bbR)$ subgroup of $\E7$
under which the $\rep{56}$ representation is given by
\begin{equation}
   E_0 = \Lambda^2V \oplus \Lambda^2V^*
\end{equation}
where $V$ is the eight-dimensional fundamental representation of
$\SL(8,\bbR)$. The $\GL(7,\bbR)$ structure group of the tangent space
$TM$ embeds as (see appendix~\ref{app:GL7})
\begin{equation}
\label{eq:Vdecomp}
   V = \big[(\Lambda^7T^*M)^{1/4}\otimes TM \big]
      \oplus (\Lambda^7T^*M)^{-3/4} .
\end{equation}
One then finds
\begin{equation}
\label{eq:E0def}
   E_0 =  (\Lambda^7T^*M)^{-1/2} \otimes \left[
         TM \oplus \Lambda^2T^*M \oplus \Lambda^5T^*M
         \oplus (T^*M \otimes\Lambda^7T^*M) \right] .
\end{equation}
(Note that the final term in brackets can also be written as
$(\Lambda^7T^*M)^2\otimes\Lambda^6TM$). The bundle
$(\Lambda^7T^*M)^{-1/2}$ is isomorphic to the trivial bundle, thus
there is always a (non-canonical) isomorphism
\begin{equation}
\label{eq:Eiso}
   E_0 \simeq TM \oplus \Lambda^2T^*M \oplus \Lambda^5T^*M
         \oplus (T^*M \otimes\Lambda^7T^*M) .
\end{equation}
It is $E_0$ which is the (untwisted) exceptional generalised tangent
space\footnote{Note that there is a second possible way to embed
  $\GL(7,\bbR)$, and hence choice for $E_0$, analogous to the choice
  of spin-structures of $O(d,d)$~\cite{Gualtieri}, where  $E_0$ is
  defined as in~\eqref{eq:E0def} except with an overall factor of
  $\Lambda^7TM/|\Lambda^7TM|$. This bundle has a similar isomorphism
  to~\eqref{eq:Eiso} but with $TM$ and $T^*M$ exchanged
  everywhere.}.
Except for the overall tensor density factor of
$(\Lambda^7T^*M)^{-1/2}$, we see that we can identify it as a sum of
vectors, two-forms, five-forms and one-forms tensor seven-forms. Given
the isomorphism~\eqref{eq:Eiso} we can write
\begin{equation}
\label{eq:Xdef}
   X = x + \omega + \sigma + \tau \in E_0 ,
\end{equation}
where, writing the $\GL(7,\bbR)$ indices explicitly, we have $x^m$,
$\omega_{mn}$,  $\sigma_{m_1\dots m_5}$ and $\tau_{m,n_1\dots
  n_7}$. Physically in M-theory we expect these to correspond to
momentum, membrane, fivebrane and Kaluza--Klein monopole charge
respectively.

Recall that the T-duality symmetry group acting on the generalised
tangent space was defined in terms of a natural $O(d,d)$-invariant
metric. As discussed in  appendix~\ref{app:E7def}, the group $\E7$ is
defined by, not a metric, but a symplectic structure $\Omega$ and
symmetric quartic invariant $q$ on the 56-dimensional representation
space. These are given explicitly in terms of $\SL(8,\bbR)$
representations in the appendix~\eqref{app:E7def}, and are by
definition $\GL(7,\bbR)$, hence diffeomorphism, invariant.

Having identified the $\GL(7,\bbR)$ tangent space symmetry in $\E7$,
we would next like to identify the analogues of the
$B$-shifts. This is essentially contained in the original dimensional
reduction of eleven-dimensional supergravity on $T^7$~\cite{CJ}. We
note that the 133-dimensional adjoint representation of $\E7$
decomposes under $\GL(7,\bbR)$ as
\begin{equation}
\label{eq:adbundle}
   A = (TM\otimes T^*M) \oplus \Lambda^6TM \oplus \Lambda^6T^*M
      \oplus \Lambda^3TM \oplus \Lambda^3T^*M .
\end{equation}
Given there is a three-form potential $A$ in eleven-dimensional
supergravity, the analogue of $B$-shifts should be $A$-shifts generated by
$A\in\Lambda^3T^*M$. In fact, we will also consider $\tA$-shifts with
$\tA\in\Lambda^6T^*M$ corresponding to the dual six-form
potential. This will be described in more detail in the next
section. For now we simply note that $A$ and $\tA$ are both elements
of the adjoint bundle~\eqref{eq:adbundle}. Their action on $X\in E$ is
given in~\eqref{eq:AA'action}. It exponentiates to
\begin{equation}
\begin{aligned}
   & \ee^{A+\tA}X =
      x + \left[\omega + i_xA\right]
      + \big[
         \sigma + A\wedge\omega + \tfrac{1}{2}A\wedge i_xA
         + i_x\tA \big] \\ &\quad
      + \big[ \tau + jA\wedge\sigma - j\tA\wedge\omega
         + jA\wedge i_x\tA
         + \tfrac{1}{2}jA\wedge A \wedge \omega
         + \tfrac{1}{6}jA\wedge A \wedge i_xA
         \big],
\end{aligned}
\end{equation}
where we are using a notion for elements of $T^*M\otimes(\Lambda^7T^*M)$
defined in~\eqref{eq:jdef}. Note that the action truncates at cubic
order. The corresponding Lie algebra, unlike the case of
$B$-shifts is not Abelian. We have the commutator
\begin{equation}
\label{eq:A-algebra}
   [ A + \tA, A' + \tA'] = - A \wedge A' .
\end{equation}
That is, two $A$-shifts commute to give a $\tA$
shift~\cite{Cremmer:1998px}.

When $A$ and $\tA$ are non-trivial one is led to defining a twisted
EGT which encodes the patching of the potentials. This is again
completely analogous to the generalised geometrical case. One starts
with
\begin{equation}
   X_0 \in TM \oplus \Lambda^2T^*M \oplus \Lambda^5T^*M
         \oplus (T^*M \otimes\Lambda^7T^*M) .
\end{equation}
On a given patch $U_{(\alpha)}$ we define the shifted element
\begin{equation}
   X_{(\alpha)} = \ee^{A_{(\alpha)} + \tA_{(\alpha)}}X_0 .
\end{equation}
In passing from one patch to another we have, on $U_{(\alpha)}\cap
U_{(\beta)}$,
\begin{equation}
\label{eq:Epatch}
   X_{(\alpha)} = \ee^{\dd\Lab+\dd\tLab}X_{(\beta)} ,
\end{equation}
provided the connections $A$ and $\tA$ patch as
\begin{equation}
\label{eq:AtApatch}
\begin{aligned}
   A_{(\alpha)} - A_{(\beta)} &= \dd\Lab , \\
   \tA_{(\alpha)} - \tA_{(\beta)} &= \dd\tLab
      - \tfrac{1}{2}\dd\Lab\wedge A_{(\beta)} .
\end{aligned}
\end{equation}
As we will see in the next section this corresponds exactly to the
patching of the three- and six-form potentials arising from
eleven-dimensional supergravity.

The patching~\eqref{eq:Epatch} imply that $X_{(\alpha)}$ are sections
of a twisted EGT which we will denote as $E$. Explicitly in components
we have
\begin{equation}
\begin{aligned}
   x_{(\alpha)} &= x_{(\beta)} , \\
   \omega_{(\alpha)} &= \omega_{(\beta)}
      + i_{x_{(\beta)}}\dd\Lab , \\
   \sigma_{(\alpha)} &= \sigma_{(\beta)}
      + \dd\Lab\wedge\omega_{(\beta)}
      + \tfrac{1}{2}\dd\Lab\wedge i_{x_{(\beta)}}\dd\Lab
      + i_{x_{(\beta)}}\dd\tLab, \\
   \tau_{(\alpha)} &= \tau_{(\beta)}
      + j\dd\Lab\wedge\sigma_{(\beta)}
      - j\dd\tLab\wedge\omega_{(\beta)}
      + j\dd\Lab\wedge i_{x_{(\beta)}}\dd\tLab \\ &\qquad {}
      + \tfrac{1}{2}j\dd\Lab\wedge\dd\Lab\wedge\omega_{(\beta)}
      + \tfrac{1}{6}j\dd\Lab\wedge\dd\Lab\wedge i_{x_{(\beta)}}\dd\Lab
\end{aligned}
\end{equation}
One can define $E$ formally via a series of extensions
\begin{equation}
\label{eq:twistE}
\begin{gathered}
   0 \longrightarrow \Lambda^2T^*M \longrightarrow E''
      \longrightarrow TM \longrightarrow 0 , \\
   0 \longrightarrow \Lambda^5T^*M \longrightarrow E'
      \longrightarrow E'' \longrightarrow 0 , \\
   0 \longrightarrow T^*M\otimes \Lambda^7T^*M
      \longrightarrow E \longrightarrow E'
      \longrightarrow 0 ,
\end{gathered}
\end{equation}
in analogy with~\eqref{eq:gen-T}.

As for the $B$-field, the potentials $A$ and $\tA$ are formally
connections on gerbes. To define the connective structure of the gerbe
we must define the patchings on successively higher-order
intersections. For $A$, on the corresponding multiple intersections of
patches we have
\begin{equation}
\begin{aligned}
   \Lambda_{(\alpha\beta)} + \Lambda_{(\beta\gamma)}
      + \Lambda_{(\gamma\alpha)} &= \dd\Lambda_{(\alpha\beta\gamma)}
      &&
      \quad \text{on $U_{(\alpha)}\cap U_{(\beta)} \cap U_{(\gamma)}$}, \\
   \Lambda_{(\beta\gamma\delta)} - \Lambda_{(\alpha\gamma\delta)}
      + \Lambda_{(\alpha\beta\delta)} - \Lambda_{(\alpha\beta\gamma)}
      &= \dd\Lambda_{(\alpha\beta\gamma\delta)}
      &&
      \quad \text{on $U_{(\alpha)}\cap U_{(\beta)} \cap U_{(\gamma)}
         \cap U_{(\delta)}$} .
\end{aligned}
\end{equation}
For a quantised flux $\FF=\dd A_{(\alpha)}$ we have
$g_{(\alpha\beta\gamma\delta)}=\ee^{\ii\Lambda_{(\alpha\beta\gamma\delta)}}\in
U(1)$ with the cocycle condition
\begin{equation}
   g_{(\beta\gamma\delta\epsilon)}g^{-1}_{(\alpha\gamma\delta\epsilon)}
      g_{(\alpha\beta\delta\epsilon)}g^{-1}_{(\alpha\beta\gamma\epsilon)}
      g_{(\alpha\beta\gamma\delta)} = 1 ,
\end{equation}
on $U_{(\alpha)}\cap \dots \cap U_{(\epsilon)}$. For $\tLab$ there is
a similar set of structures, with the final cocycle condition defined
on a octuple intersection $U_{(\alpha_1)}\cap\dots\cap
U_{(\alpha_8)}$.

The bundle $E$ encodes all the topological information of the
supergravity background: the twisting of the tangent space $TM$ as
well as the patching of the form potentials, but, as
for~\eqref{eq:gen-T} is independent of the particular choice of $A$
and $\tA$.

Finally we would like to identify the analogue of the Courant bracket
for the EGT. We look for a pairing with $A$- and $\tA$-shifts as
automorphisms when $\dd A=\dd\tA=0$. One finds the unique
``exceptional Courant bracket'' (ECB):
\begin{equation}
\label{eq:ECB}
\begin{split}
   \big[ x+&\omega+\sigma+\tau, x'+\omega'+\sigma'+\tau' \big] = \\
     & [x,x']
        + \mathcal{L}_x\omega' - \mathcal{L}_{x'}\omega
           - \tfrac{1}{2}\dd\left(i_x\omega'-i_{x'}\omega\right)
     \\ & \quad
        + \mathcal{L}_x\sigma' - \mathcal{L}_{x'}\sigma
           - \tfrac{1}{2}\dd\left(i_x\sigma'-i_{x'}\sigma\right)
        + \tfrac{1}{2}\omega\wedge\dd\omega'
           - \tfrac{1}{2}\omega'\wedge\dd\omega
     \\ & \quad
        + \tfrac{1}{2}\mathcal{L}_x\tau'
           - \tfrac{1}{2}\mathcal{L}_{x'}\tau
        + \tfrac{1}{2}\big(
           j\omega\wedge\dd\sigma' - j\sigma'\wedge\dd\omega \big)
        - \tfrac{1}{2}\big(
           j\omega'\wedge\dd\sigma - j\sigma\wedge\dd\omega' \big)
\end{split}
\end{equation}
If $\SG$ is the group generated by closed $A$ and $\tA$
shifts, the ECB is invariant under the $\Diff(M)\ltimes\SG$.


\subsection{The exceptional generalised metric and $\hg$ structures}
\label{sec:EGM}

Having defined the EGT, its topology and the corresponding bracket,
one can then introduce, as in~\cite{chris}, the analog of the
generalised metric which encodes the fields of eleven-dimensional
supergravity. The motivation is that, when compactified on $T^7$, the
moduli arising from the eleven-dimensional supergravity fields $g$ and
$A$ parametrise a $\E7/(\hg)$ coset space~\cite{CJ}. Rather than
consider a fixed element of the coset space, one takes one that is a
function of position in the manifold $M$~\cite{deWN}. In the
following, instead of starting with the coset, we show how the
exceptional generalised metric can be defined as a generic $\hg$
structure on the EGT.

Fixing an element of the coset space $\E7/(\hg)$ is equivalent to
choosing a particular $\hg$ subgroup of $\E7$. Making such a choice at
each point in $M$ corresponds geometrically to a $\hg$ structure on
$E$. Such a structure can be defined as follows. The $\rep{56}$
representation decomposes into $\rep{28}+\bar{\rep{28}}$ under
$\hg$. Thus an $\hg$ structure is equivalent to the existence of a
decomposition of the (complexified) EGT $E\otimes\bbC=C\oplus\bar{C}$
where the fibres of $C$ transform in the $\rep{28}$ of $\hg$. However
this is the same as an almost complex structure $J$ with $J^2=-\id$ on
$E$. For $J$ to define an $\hg$ subgroup it must also be compatible
with the $\E7$ structure. Recall that the latter was defined by a
symplectic form $\Omega$ and a quartic invariant $q$. Compatibility
requires
\begin{equation}
   \Omega(JX,JY) = \Omega(X,Y) , \qquad
   q(JX) = q(X) ,
\end{equation}
or in other words $J\in\E7$. (Note the that first condition is just
the usual condition between a symplectic and an almost complex
structure required to define an Hermitian metric.) Such an almost
complex structure $J$ defines an $\hg$ structure on $E$.

In contrast to the generalised geometry case where the $O(d)\times
O(d)$ structure was equivalent to a compatible almost product
structure satisfying $\Pi^2=\id$, for $\hg\subset\E7$ the structure is
defined by a compatible almost complex structure satisfying
$J^2=-\id$. Given $J$ and $\Omega$ one can then define the
corresponding exceptional generalised metric (EGM) $G$ by
\begin{equation}
   G(X,Y) = \Omega(X,JY) ,
\end{equation}
which gives a positive definite metric on $E$.

We now turn to how one constructs the generic form of $J$ and hence
$G$. Given a metric $\tg_{ab}$ on the $\SL(8,\bbR)$ representation
space $V$, a natural way to define a particular almost complex
structure $J_0$ (using the conventions of appendix~\ref{app:E7def}, so
that, in particular, pairs of indices $aa'$ and $bb'$ are
antisymmetrised) is as
\begin{equation}
   J_0X = \begin{pmatrix}
              0 & - \tg^{ab}\tg^{a'b'} \\
              \tg_{ab}\tg_{a'b'} & 0
           \end{pmatrix}
           \begin{pmatrix} x^{bb'} \\ x'_{bb'} \end{pmatrix}
       = \begin{pmatrix}
              - \tg^{ab}\tg^{a'b'}x'_{bb'} \\ \tg_{ab}\tg_{a'b'}x^{bb'}
          \end{pmatrix} .
\end{equation}
By construction $J_0^2=-\id$. The corresponding EGM is
\begin{equation}
      G_0(X,Y) = \tg_{ab}\tg_{a'b'}x^{aa'}y^{bb'}
         + \tg^{ab}\tg^{a'b'}x'_{aa'}y'_{bb'} .
\end{equation}
From the definitions~\eqref{eq:Odef} and~\eqref{eq:qdef} of $\Omega$
and $q$ it is clear that $J_0\in\E7$ provided
$\det\tg=1$. Under an infinitesimal $\E7$ transformation
$\mu\in\rep{133}$ we have
\begin{equation}
   \delta J_0 = [ \mu , J_0 ]
      = \begin{pmatrix}
         \mu^{+aa'}{}_{bb'} & - 2 \mu^{+ab}\tg^{a'b'} \\
         - 2 \mu^+_{ab}\tg_{a'b'} & - \mu^+_{aa'}{}^{bb'}
      \end{pmatrix}
\end{equation}
where $\mu^\pm_{abcd}=\tfrac{1}{2}(\mu_{abcd}\pm *\mu_{abcd})$ and
$\mu^\pm_{ab}=\tfrac{1}{2}(\mu_{ab}\pm\mu_{ba})$ where indices are raised and lowered
using $\tg$. Thus $J_0$ is invariant under the subgroup generated by
$\mu^-_{ab}$ and $\mu^-_{abcd}$. As discussed in
appendix~\ref{app:SU8} this is precisely $\hg$ (see~\eqref{eq:SU8decomp}).

Given the embedding~\eqref{eq:Vdecomp} of
$\GL(7,\bbR)\subset\SL(8,\bbR)$ discussed in detail in
appendix~\ref{app:GL7}, we can define $\tg$ in terms of a
seven-dimensional metric $g$ as
\begin{equation}
\label{eq:tgdef}
   \tg_{ab} = (\det g)^{-1/4} \begin{pmatrix}
         g_{mn} & 0 \\
         0 & \det g
      \end{pmatrix} .
\end{equation}
Acting on elements of $X=x+\omega+\sigma+\tau$ we have
\begin{equation}
   G_0(X,X) = 2\left(|x|^2 + |\omega|^2 + |\sigma|^2 + |\tau|^2 \right) ,
\end{equation}
where $|\tau|^2=\frac{1}{7!}\tau_{m.n_1\dots n_7}
\tau^{m,n_1\dots n_7}$, $|\sigma|^2=\frac{1}{5!}\sigma_{n_1\dots
  n_5}\sigma^{n_1\dots n_5}$ etc. and, so that the result is a scalar, we
have dropped on overall factor of $(\det g)^{1/2}$, which is natural,
since in writing $X=x+\omega+\sigma+\tau$ we are using the
isomorphism~\eqref{eq:Eiso}.

Given a seven-dimensional metric $g$ we have been able to write a
particular $\hg$ structure $J_0$. A generic structure, given all such
structures lie in the same orbit, will be of the form $J=hJ_0h^{-1}$ where
$h\in\E7$, or equivalently $G(X,Y)=G_0(h^{-1}X,h^{-1}Y)$. We write
$h=\ee^\mu$ with the Lie algebra element
$\mu=(\mu^a{}_b,\mu_{abcd})\in\rep{133}$. The elements $\mu^m{}_n$
generate the $\GL(7,\bbR)$ subgroup and acting on $J_0$ simply change
the form of the metric $g$. The additional components $\mu^8{}_m$ and
$\mu^m{}_8$ modify the form of $\tg$~\eqref{eq:tgdef}. Since only
$\mu^+_{ab}$ acts non-trivially on $J_0$, we need only consider
transformations with, say, $\mu^m{}_8$. Similarly since only
$\mu^+_{abcd}$ acts non-trivially we can generate a generic $J$ using
only, say $\mu_{mnp8}$. However, $\mu^m{}_8$ and $\mu_{mnp8}$
transformations precisely correspond to the subgroup of $A$- and
$\tA$-shifts. Thus, given a generic $g$ defining $G_0$, the generic
EGM can be written as
\begin{equation}
   G(X,Y) = G_0(\ee^{-A-\tA}X,\ee^{-A-\tA}Y) .
\end{equation}
This is analogous to the form~\eqref{eq:B-shiftG} of the generalised
metric in generalised geometry. Note also that for non-trivial $A$ and
$\tA$, $G_0$ is an EGM on the untwisted EGT given by
$TM\oplus\Lambda^2T^*M\oplus\Lambda^5T^*M
\oplus(T^*M\otimes\Lambda^7T^*M$), while $G$ is an EGM on the twisted
bundle $E$ given by~\eqref{eq:twistE}.


\section{Supersymmetric backgrounds and EGG}
\label{sec:SEGG}


We will now relate the EGG defined in the previous section
to eleven-dimensional supergravity and in particular seven-dimensional
supersymmetric backgrounds. After identifying the standard
decomposition of the supergravity degrees of freedom on such
backgrounds, we first briefly review the corresponding connection
between type II backgrounds and generalised geometry before turning to
EGG.

The physical context we are interested in is where the
eleven-dimensional spacetime is topologically a product of a
four-dimensional ``external'' and a seven-dimensional ``internal''
space
\begin{equation}
\label{eq:M11}
   M_{10,1} = M_{3,1} \times M_7
\end{equation}
If $M_7$ is compact we can consider compactifying eleven-dimensional
supergravity to give an effective four-dimensional theory. In
particular, the effective theory could be supersymmetric. We could
also look for particular examples of compactifications which are
solutions of the supergravity field equations and preserve some number
of supersymmetries. In either case, the geometry of $M_7$ is
restricted, and some discussion of the latter case appears
in~\cite{chris}. The goal here is to understand how this restricted
geometry can be naturally described in terms of EGG structures on
$M_7$. We will also focus on the low-energy effective theory rather
than the on-shell supersymmetric backgrounds.


\subsection{Effective theories and field decompositions}
\label{sec:fdecomp}

Given the product~\eqref{eq:M11}, the tangent bundle decomposes as
$TM_{10,1}=TM_{3,1}\oplus TM_7$ and all the supergravity fields can be
decomposed under a local $\Spin(3,1)\times\Spin(7)\subset\Spin(10,1)$
symmetry. Normally one would derive a four-dimensional effective
description by truncating the Kaluza--Klein spectrum of modes on $M_7$
to give a four-dimensional theory with a finite number of degrees of
freedom. For instance, compactifying on a torus and keeping massless
modes, one finds that the degrees of freedom actually arrange
themselves into multiplets transforming under $\E7$ for the bosons and
$\SU(8)$ for the fermions.

However, one can also keep the full dependence of all
eleven-dimensional fields on both the position on $M_{3,1}$ and
$M_7$. One can then simply rewrite the eleven-dimensional theory,
breaking the local $\Spin(10,1)$ symmetry to
$\Spin(3,1)\times\Spin(7)$, so that it is analogous to a
four-dimensional theory. This was done explicitly by de Wit and
Nicolai~\cite{deWN}, retaining all 32 supersymmetries, where it was
shown that in general the degrees of freedom fall into $\E7$ and $\hg$
representations. In this paper we will ultimately be interested in
such reformulations focusing on only four of the supercharges so that
the theory has a structure analogous to $N=1$ four-dimensional
supergravity. Note that formally the only requirement for making such
rewritings is not that $M_{10,1}$ is topologically a product, but
rather that the tangent space $TM_{10,1}$ decomposes into a four- and
seven-dimensional part
\begin{equation}
   TM_{10,1} = T \oplus F .
\end{equation}
For simplicity, here we will concentrate on the case of a product
manifold, though all of our analysis actually goes through in the more
general case, with the EGT defined in terms of $F$ rather than
$TM_7$. The analogous analysis in terms of generalised geometry for
type II compactifications was given in~\cite{GLW1,GLW2}.

Let us briefly note how the fields decompose under
$\Spin(3,1)\times\Spin(7)$. Our conventions for eleven-dimensional
supergravity are summarised in appendix~\ref{app:sugra}. The degrees
of freedom are the metric $g_{MN}$, three-form $A_{MNP}$ and gravitino
$\Psi_M$. Consider first the $\Spin(3,1)$ scalars. The
eleven-dimensional metric decomposes as a warped product
\begin{equation}
   \dd s^2(M_{11}) = \ee^{2E}g^{(4)}_{\mu\nu}\dd x^\mu\dd x^\nu
      + g_{mn} \dd x^m \dd x^n ,
\end{equation}
where $\mu=0,1,2,3$ denote coordinates on the external space. To get a
conventionally normalised Einstein term in the four-dimensional
effective theory we must take
\begin{equation}
   \ee^{-2E} = \sqrt{\det g} ,
\end{equation}
where $\det g$ is the determinant of $g_{mn}$. In a conventional compactification, deformations of the internal
metric $g_{mn}$ lead to scalar moduli fields in the effective
theory. Moduli fields can also arise from the flux $F$. Keeping only
$\Spin(3,1)$ scalar parts, one can decompose
\begin{equation}
   F = *_7\tFF \wedge \ee^{4E}\epsilon_{(4)} + \FF
\end{equation}
where $\FF\in\Lambda^4T^*M_7$ and $\tFF\in\Lambda^7T^*M_7$ and
$\epsilon_{(4)}=\sqrt{-g^{(4)}}\dd x^0\wedge\dots\wedge\dd x^3$. The
eleven-dimensional equation of motion and Bianchi
identify~\eqref{eq:Feom} then decompose as
\begin{equation}
\begin{aligned}
   \dd \tFF + \tfrac{1}{2}\FF\wedge \FF &= 0 , & \qquad \qquad
   \dd \big(\ee^{4E}*_7 \tFF\big) &= 0 , \\
   \dd \FF &= 0 , & \qquad \qquad
   \dd \big(\ee^{4E}*_7 \FF\big)
       + \ee^{4E}*_7\tFF\wedge\FF &= 0
\end{aligned}
\end{equation}
so one can introduce, locally,
\begin{equation}
\begin{aligned}
   \FF & = \dd A , \\
   \tFF &= \dd\tA - \tfrac{1}{2}A\wedge \FF ,
\end{aligned}
\end{equation}
where $A\in\Lambda^3T^*M_7$ and $\tA\in\Lambda^6T^*M_7$. By
definition, $\FF$ and $\tFF$ are globally elements of $\Lambda^4T^*M$
and $\Lambda^7T^*M$ respectively. Thus on $U_{(\alpha)}\cap
U_{(\beta)}$ we have
\begin{equation}
\begin{aligned}
   \dd A_{(\alpha)} - \dd A_{(\beta)} &= 0 \\
   \dd \tA_{(\alpha)} - \dd \tA_{(\beta)}
      &= \tfrac{1}{2}\left(A_{(\alpha)}-A_{(\beta)}\right)
         \wedge\dd A_{(\beta)}
\end{aligned}
\end{equation}
This implies that the potentials $A$ and $\tA$ must patch precisely as
given by~\eqref{eq:AtApatch}. We see that the twisting of the
EGT~\eqref{eq:Epatch} is precisely that corresponding to the
supergravity potentials. Furthermore, given the discussion of the
previous section~\ref{sec:EGM} the scalar degrees of freedom $g_{mn}$
and $A_{mnp}$ and $\tA_{m_1\dots m_6}$ scalars can be combined
together as an EGM or equivalently an almost complex structure $J$ on
$E$.

Turning briefly to the remaining fields, there are 28 bosonic
$\Spin(3,1)$ vector degrees of freedom coming from off-diagonal
components of the metric $g_{\mu m}$ and from $A_{\mu mn}$. One
usually also introduces the corresponding dual potentials giving a
total of $56$. Finally for the fermionic degrees of freedom we
decompose the eleven-dimensional gamma matrices as
\begin{equation}
\label{gamma-decomp}
   \Gamma^\mu = \ee^{-E} \gamma^\mu \otimes \id \qquad
   \Gamma^m = \gfive \otimes \gamma^m .
\end{equation}
The seven-dimensional gamma-matrix conventions are defined
in~\ref{app:7d}, while the four-dimensional gamma matrices are
chosen to satisfy $\{\gamma_\mu,\gamma_\nu\}=2g^{(4)}_{\mu\nu}\id$ and
$\gamma_{\mu_1\dots\mu_4}=\gamma_{(4)}\epsilon^{(4)}_{\mu_1\dots\mu_4}$. The
real eleven-dimensional spinors correspondingly decompose as
\begin{equation}
\label{eq:spinor-decomp}
\begin{aligned}
   \epsilon &= \ee^{E/2}\theta_+\otimes{\ksp^c} +
      \ee^{E/2}\theta_-\otimes\ksp \\
   \rep{32} &= (\rep{2},\rep{8}) + (\bar{\rep{2}},\bar{\rep{8}})
\end{aligned}
\end{equation}
where $\gfive\theta_\pm=\pm\theta_\pm$ (with
$\theta_+^c=D\theta^*_+=\theta_-$ and $-\gamma_\mu^*=D^{-1}\gamma_\mu
D$) are chiral four-dimensional spinors and ${\ksp^c}$ is a complex
$\Spin(7)$ spinor. The factor of $\ee^{E/2}$ and the choice of
labelling $\ksp$ versus $\ksp^c$ are conventional. Thus $\Psi_\mu$
decomposes as eight spin-$\frac{3}{2}$ fermions, while $\Psi_m$ gives
56 spin-$\frac{1}{2}$ fermions.

As discussed in appendix~\ref{app:7d} there is a natural embedding of
$\SU(8)$ in the Clifford algebra $\Cliff(7,0;\bbR)$ with the complex
$\Spin(7)$ spinors transforming in the fundamental representation. In
reformulating the eleven-dimensional theory, all the degrees of
freedom, fermionic and bosonic arrange as $\SU(8)$
representations~\cite{deWN}. Thus we can actually promote the
$\Spin(3,1)\times\Spin(7)$ symmetry to $\Spin(3,1)\times\SU(8)$. This
decomposition corresponds to the $N=8$ four-dimensional supergravity
multiplet. It is summarized in table~\ref{tab:N=8} where
$\rep{r_s}$ transforms  as the $\rep{r}$ representation of $\SU(8)$
with $\Spin(3,1)$ spin $\rep{s}$.
\TABLE[h]{
   \begin{tabular}{rlcrl}
      $g_{mn}$, $A_{mnp}$, $\tA_{m_1\dots m_6}$ : &
         $\rep{35_0}+\bar{\rep{35}}_{\rep{0}}$
      \qquad \qquad &&
      $\Psi_m$ : & $\rep{56_{1/2}}$ \\
      $g_{\mu m}$, $A_{\mu mn}$ + duals : &
         $\rep{28_1}+\bar{\rep{28}}_{\rep{1}}$ &&
      $\Psi_\mu$ : & $\rep{8_{3/2}}$ \\
      $g^{(4)}_{\mu\nu}$ : & $\rep{1_2}$
   \end{tabular}
   \caption{Decomposition of eleven-dimensional supergravity fields
     under $\Spin(3,1)\times\SU(8)$}
   \label{tab:N=8}
}

To summarise, from a EGG perspective, the scalar degrees of freedom
define an $\hg$ structure on the EGT. Given this structure (or rather
the existence of a double cover $\SU(8)$ structure, which is not
always guaranteed) one can then define $\SU(8)$ spinors and hence the
fermionic degrees of freedom. This is the EGG analogue of requiring a
metric, or $O(d)$ structure, and hence a set of vielbeins, before one
can define ordinary spinors on a curved manifold.


\subsection{Review of generalized geometry of $N=2$ type II backgrounds}

The generic effective four-dimensional $N=2$ supersymmetric theories
arising from type II supergravity compactified on $M_{3,1}\times M_6$
were analysed in terms of generalised geometry
in~\cite{GLW1,GLW2}. The structure was as follows.

The metric and $B$-field on $M_6$ combine into a generalised
metric~\eqref{eq:gen-G}. This defines a $O(6)\times O(6)$ structure on
the generalised tangent space~\eqref{eq:gen-T}, that is the
decomposition $E=C^+\oplus C^-$. Assuming the double cover
$\Spin(6)\times\Spin(6)$ exists, one can define $\Spin(6)$ spinors on
$C^+$ and $C^-$ separately. In terms of the original ten-dimensional
spinors one has the decomposition
\begin{equation}
\begin{aligned}
   \epsilon^1 &= \theta^1_+ \otimes \ksp^1_-
      + \theta^1_- \otimes \ksp^1_+ \\
   \epsilon^2 &= \theta^2_+ \otimes \ksp^2_\pm
      + \theta^2_- \otimes \ksp^2_\mp
\end{aligned}
\end{equation}
where in the second line one takes the upper sign for type IIA and the
lower for type IIB. (Here $\ksp_\pm$ are complex, chiral $\Spin(6)$
spinors, with $\ksp_-=\ksp_+^c$.) The two spinors $\ksp^1_+$ and
$\ksp^2_+$ naturally transform under the two spin groups
$\Spin(6)\times\Spin(6)$.

Next, we want to concentrate on effective theories with $N=2$
supersymmetry in four dimensions. This means, we want to identify a
fixed pair of spinors $(\ksp^1_+,\ksp^2_+)$. The eight
four-dimensional supersymmetric parameters are then parametrised by
$(\theta^1_+,\theta^2_+)$. In order to be able to use these
supersymmetries to decompose all modes of the ten-dimensional fields
into $N=2$ multiplets, we must require that $(\ksp^1_+,\ksp^2_+)$ are
non-vanishing and globally defined. (We can then project the
supersymmetry from an action on $M_{9,1}$ to an action on $M_{3,1}$.)
But this condition is the same as requiring each spinor to define an
$\SU(3)$ structure. Thus we have
\begin{equation}
   \text{$N=2$ effective theory} \Leftrightarrow
   \text{$\stt$ structure on $E$} .
\end{equation}
The structure can be defined by the existence of the generalised
metric $G$ on $E$ together with a pair of $\Spin(6)\times\Spin(6)$
spinors $(\ksp^1_+,\ksp^2_+)$.

A conventional $\SU(3)$ structure can similarly be defined by a
ordinary metric $g$ together with a globally defined, nowhere
vanishing spinor $\ksp_+$. However, one can also define the structure
by a pair of real forms $J\in\Lambda^2T^*M_6$ and
$\rho\in\Lambda^3T^*M_6$. In the case where the $\SU(3)$ structure is
integrable, that is when $M_6$ is a Calabi--Yau manifold, $J$ is the
K\"ahler form and $\rho$ is the real part of the holomorphic
three-form\footnote{Note that $\hat{\rho}$ can be determined as a
  homogeneous function of $\rho$ of degree one.}
$\Omega=\rho+\ii\hat{\rho}$. Generically $(J,\rho)$ only define a
$\SU(3)$ structure if  $J\wedge\rho=0$ and $J\wedge J\wedge
J=\frac{3}{2}\rho\wedge\hat{\rho}$.

It is natural to ask if the $\stt$ structure can be similarly defined
in terms of $O(6,6)$ objects. It can, and the representations in
question are the spinors of $\Spin(6,6)$. These are defined as
follows. (For more details see appendix~A of~\cite{GLW2}.) For
$E_0=TM_6\oplus T^*M_6$  the spinor bundle $S(E_0)$ is isomorphic to
the bundle of forms
\begin{equation}
\label{spinbundles}
  S(E) = (\Lambda^7T^*M_6)^{-1/2} \otimes \Lambda^* T^* M_6 .
\end{equation}
More generally, for $E$, an extension of the form~\eqref{eq:gen-T}, on
any patch $U_{(\alpha)}$, a spinor $\Upsilon_{(\alpha)}\in S(E)$ is a
sum of forms, with the patching
\begin{equation}
   \Upsilon_{(\alpha)} = \ee^{-\dd\Lambda_{(\alpha\beta)}}\Upsilon_{(\beta)} ,
\end{equation}
where the action is by wedge product. Spinors of $O(6,6)$ are
Majorana--Weyl. The positive and negative helicity spin bundles
$S^\pm(E)$ are locally isomorphic to the bundles of even and odd forms
$\Leo T^*M_6$. The Clifford action on $\Upsilon\in S(E)$, viewed as a
sum of forms, is given by
\begin{equation}
\label{cliff}
   (x+\xi)\cdot\Upsilon = i_x\Upsilon + \xi\wedge\Upsilon .
\end{equation}
The usual spinor bilinear form on $S(E)$ is given by the Mukai pairing
$\mukai{\cdot}{\cdot}$ on forms. Explicitly
\begin{equation}
\label{eq:mukai}
   \mukai{\Upsilon}{\Upsilon'}
     = \sum_p (-)^{[(p+1)/2]} \Upsilon_{(p)} \wedge \Upsilon'_{(6-p)}\ ,
\end{equation}
where the subscripts denote the degree of the component forms in
$\Lambda^*T^*M_6$ and $[(p+1)/2]$ takes the integer part of
$(p+1)/2$. Note that, given the isomorphism~\eqref{spinbundles}, the
ordinary exterior derivative defines a natural generalised Dirac
operator $\dd:S^\pm(E)\to S^\mp(E)$.

Given a generalised metric $G$, on can decompose $\Spin(6,6)$ spinors
under $\Spin(6)\times\Spin(6)$. Projecting each subspace $C^\pm$ onto
$TM$ defines a common $\Spin(6)$ subgroup. Under this group $\Upsilon\in
S(E)$ transform as a bispinor, that is, as an element of
$\Cliff(6,0;\bbR)$. Explicitly one can write real $\Upsilon^\pm\in
S^\pm$ as
\begin{equation}
\label{66decomp}
   \Upsilon^\pm = \zeta_+\bar{\zeta}'_\pm
      \pm  \zeta_-\bar{\zeta}'_\mp\ ,
\end{equation}
where $\zeta_+$, $\zeta'_+$ are ordinary $\Spin(6)$ spinors and
elements of the $\Spin(d)$ bundles $S^+(C^+)$ and $S^+(C^-)$
respectively. From this perspective $\Upsilon^\pm$ is a matrix. It can
be expanded as
\begin{equation}
\label{chichiepsilon}
   \Upsilon^\pm = \sum_p
      \frac{1}{8p!}\Upsilon^\pm_{m_1\dots m_p} \gamma^{m_1\dots m_p} ,
\end{equation}
with
\begin{equation}
   \Upsilon^\pm_{m_1\dots m_p} = \tr (\Upsilon^\pm \gamma_{m_p\dots m_1})
      \in \Lambda^pT^*M_6 ,
\end{equation}
and where $\gamma^m$ are $\Spin(6)$ gamma-matrices and the trace is
over the $\Spin(6)$ indices. For $\Upsilon^+$ only the even forms
are non-zero, while for $\Upsilon^-$ the odd forms are
non-zero. This gives an explicit realisation of the isomorphism
between $S^\pm(E)$ and $\Leo T^*M_6$.

One can introduce a pair of $\Spin(6)\times\Spin(6)$ spinors which
define the $\stt$ structure. Write the complex objects
\begin{equation}
\label{purespinors}
   \Phi^+ := \ee^{-B} \ksp^1_+\bar{\ksp}^2_+ , \qquad
   \Phi^- := \ee^{-B} \ksp^1_+\bar{\ksp}^2_- ,
\end{equation}
where again $\ee^{-B}$ acts by wedge product. Note that in the special
case where $\ksp^1_+=\ksp^2_+$, the two $\SU(3)$ structures are the
same, and we have
\begin{equation}
   \Phi^+ = \tfrac{1}{8} \ee^{-B-\ii J} , \qquad
   \Phi^- = - \tfrac{\ii}{8}\ee^{-B}\Omega .
\end{equation}
Generically, each $\Phi^\pm$ individually defines an $\SU(3,3)$
structure on $E$. Provided these structures are compatible, together
they define a common $\SU(3)\times\SU(3)$ structure. The requirements
of compatibility, in terms of the Mukai pairing~\eqref{eq:mukai} is
that~\cite{GLW1}
\begin{equation}
\label{eq:compat}
\begin{aligned}
   \mukai{\Phi^+}{V\cdot\Phi^-} &=  \mukai{\bar
     \Phi^+}{V\cdot\Phi^-}=0  \quad \forall \, V\in E , \\
   \mukai{\Phi^+}{\bar{\Phi}^+} &= \mukai{\Phi^-}{\bar{\Phi}^-} .
\end{aligned}
\end{equation}
If $\Phi^\pm$ are given by~\eqref{purespinors} they are automatically
compatible~\cite{GMPT2}. However, one can also reverse the logic. The
$\SU(3,3)$ structures can actually be defined using only the real
parts $\Upsilon^\pm=\re\Phi^\pm$. Furthermore any pair of $\Spin(6,6)$
spinors $(\Upsilon^+,\Upsilon^-)$ satisfying the conditions~\eqref{eq:compat}
define an $\stt$ structure. So
\begin{equation}
   \text{$\stt$ structure on $E$} \Leftrightarrow
      (\Upsilon^+,\Upsilon^-) .
\end{equation}

The spinor bundles $S^\pm(E)$ are 32-dimensional. The compatibility
requirement~\eqref{eq:compat} gives 13 conditions. Thus the space of
$(\Upsilon^+\Upsilon^-)$ is 51-dimensional. Different structures
$(\ksp^+,\ksp^-)$ can be related by $O(6,6)$ transformations. Since a
given structure is $\stt$ invariant, the space of all structures
should be related to the coset space $\Sigma=O(6,6)/\stt$, which is
50-dimensional. From this perspective the compatible pair
$(\Upsilon^+,\Upsilon^-)$ give an embedding, as a one-dimensional
family of orbits,
\begin{equation}
   (\Upsilon^+, \Upsilon^-) : \frac{O(6,6)}{\stt} \times \bbR^+
      \hookrightarrow S^+(E)\oplus S^-(E) .
\end{equation}
The additional $\bbR^+$ factor corresponds simply to an overall
rescaling of the generalised spinors.

The key point of introducing all these structures is that they provide
a very simple way to characterise the effective theory. As shown
in~\cite{GLW1,GLW2}, $\Phi^\pm$ parametrise a special K\"ahler space
underlying the vector or hypermultiplet scalar degrees of freedom, and
there are simple expressions for the $N=2$ analogues of the
superpotential in terms of $\Phi^\pm$.


\subsection{$N=1$ M-theory backgrounds, EGG and $\SU(7)$ structures}

We would now like to identify the analogue of the $N=2$ $\stt$
structure of type II theories for $N=1$ compactifications of
eleven-dimensional supergravity. An $\stt$ structure is the
generalised geometrical extension of a conventional supersymmetric
$\SU(3)$ structure and hence, when on-shell, generalises the notion of
a Calabi--Yau three-fold. The structure introduced here is similarly the
EGG generalisation of a $G_2$ structure on a seven-dimensional
manifold.

Identifying an $N=1$ background requires picking out four preferred
supersymmetries out of 32, or equivalently a fixed seven-dimensional
spinor ${\ksp}$ in the general
decomposition~\eqref{eq:spinor-decomp}. This decomposition is the
most general compatible with four-dimensional Lorentz
invariance~\cite{LS}, and generically defines a complex ${\ksp}$ on
the internal space. Given an EGM $G$ we have an $\hg$ structure on
$E$, and ${\ksp}$ transforms as the fundamental representation
$\rep{8}$ of double cover $\SU(8)$. To define a generic low-energy
effective theory the spinor ${\ksp}$ must be globally defined and
nowhere vanishing. The stabilizer group in $\SU(8)$ of a fixed element
of the vector $\rep{8}$ representation is $\SU(7)$. Thus, given a
fixed spinor $\ksp$ at each point of $M$, we see that
\begin{equation}
   \text{$N=1$ effective theory} \Leftrightarrow
      \text{$\SU(7)$ structure on $E$} .
\end{equation}
The projection $E\to TM_7$ defines a $\GL(7,\bbR)$ subgroup of
$\E7$. Given a EGM $G$, this defines a $\Spin(7)\subset\SU(8)$
subgroup. We can then decompose $\ksp$ into real $\Spin(7)$ spinors,
\begin{equation}
   {\ksp} = \ksp_1 + \ii\ksp_2
\end{equation}
Each $\ksp_i$ is stabilised by a $G_2\in\Spin(7)$
subgroup. Thus from the point of the view of the ordinary tangent
space $TM_7$, if $\ksp_i$ are globally defined and non-vanishing we
have a pair of $G_2$ structures. However, all we really require is
a globally defined non-vanishing complex ${\ksp}$. Thus in general we
may not have either $G_2$ structure. In analogy to the case where
$\ksp$ is real and we have a single $G_2$ structure, we can define the
complex bilinears
\begin{equation}
\label{eq:bilins}
   \varphi_{mnp} = \ii\bar{\ksp}^c\gamma_{mnp}\ksp
   \quad \text{and} \quad
   \psi_{mnpq} = (*_7\varphi)_{mnpq} = -\bar{\ksp}^c\gamma_{mnpq}\ksp .
\end{equation}
Locally, the pair of $\ksp_i$ are preserved by a $\SU(3)$ group.

We have seen that one way to define the structure is as the pair of
EGM and $\SU(8)$ spinor $(G,{\ksp})$. However, as in the type II case,
we can also find an element lying in a particular orbit in an $\E7$
representation which can also be used to define the structure. We
expect that it can be defined as a spinor bilinear. As we discuss in a
moment, this space should also correspond to the $N=1$ chiral
multiplet space in the four-dimensional effective theory.

Decomposing under $\SU(7)$, the $\rep{56}$ representation has no
singlets so cannot have elements stabilized by $\SU(7)$. The adjoint
$\rep{133}$ does have a singlet. In terms of the spinor ${\ksp}$, the
singlet in $\mu\in\rep{133}$, using its decomposition
$\rep{133}=\rep{63}+\rep{35}+\bar{\rep{35}}$ under $\SU(8)$, can be
written as
\begin{equation}
\begin{aligned}
   \mu_0 &= \big( {\mu_0}^\alpha{}_\beta,
            \mu_{0\;\alpha\beta\gamma\delta},
            \bar{\mu}_0{}^{\alpha\beta\gamma\delta} \big) , \\
       &= \big(
            {\ksp}^\alpha\bar{{\ksp}}_\beta
               -
               \tfrac{1}{8}(\bar{{\ksp}}{\ksp})\delta^\alpha{}_\beta,
               0, 0 \big) .
\end{aligned}
\end{equation}
However, it is easy to see that this is stabilized by $U(7)$ rather
than $\SU(7)$. The next smallest $\E7$ representation is
$\rep{912}$. (See appendix~\ref{app:reps} for our conventions for
$\E7$ representations.) We can define the following $\SU(7)$-singlet
complex element in terms of its $\SU(8)$ decomposition, that is
$\rep{912}=\rep{36}+\rep{420}+\bar{\rep{36}}+\bar{\rep{420}}$,
\begin{equation}
\label{eq:phi0}
\begin{aligned}
   \phi_0 &= (\phi_0{}^{\alpha\beta},\phi_0{}^{\alpha\beta\gamma}{}_\delta,
             \bar{\phi}_{0\;\alpha\beta},
             \bar{\phi}_{0\;\alpha\beta\gamma}{}^\delta ) \\
        &= ( \ksp^\alpha\ksp^\beta, 0 , 0 , 0 ) ,
\end{aligned}
\end{equation}
Finally we can form the structure together with the form-field
potentials
\begin{equation}
   \phi = \ee^{A+\tA}\phi_0 .
\end{equation}
Such a $\phi$ does indeed define a generic $\SU(7)$ structure.

To see this, we first note that under an infinitesimal $\E7$
transformation we have
\begin{equation}
\begin{aligned}
   \delta\phi_0{}^{\alpha\beta} &= (\mu^\alpha{}_\gamma\ksp^\gamma)\ksp^\beta
      + \ksp^\alpha(\mu^\beta{}_\gamma\ksp^\gamma) , \\
   \delta\phi_0{}^{\alpha\beta\gamma}{}_\delta &= 0 , \\
   \delta\bar{\phi}_{0\;\alpha\beta} &= 0 ,\\
   \delta\bar{\phi}_{0\;\alpha\beta\gamma}{}^\delta
      &= \mu_{\alpha\beta\gamma\epsilon}\ksp^\epsilon\ksp^\delta .
\end{aligned}
\end{equation}
Thus $\phi_0$ is stabilized by elements of $\E7$ satisfying
\begin{equation}
   \mu^\alpha{}_\beta \ksp^\beta = 0 , \qquad
   \mu_{\alpha\beta\gamma\delta}\ksp^\delta = 0 .
\end{equation}
Since $\bar{\mu}_{\alpha\beta\gamma\delta}=
*\mu_{\alpha\beta\gamma\delta}$, the second condition can only be
satisfied if $\mu_{\alpha\beta\gamma\delta}=0$. Since
$\mu^\alpha{}_\beta$ is an element of the adjoint of $\SU(8)$ we see
that $\phi_0$ is indeed stabilised by $\SU(7)$. Since
$\ee^{A+\tA}\in\E7$ the stabilizer of $\phi$ must also be
$\SU(7)$. Finally, note that the $\SU(8)$ representations were defined
using the gamma matrices $\tgam^a$ defined using the seven dimensional
metric $g$. Since action by $\ee^{A+\tA}$ generates a generic EGM,
we see that (when taken with the choice of generic $g$ and spinor
${\ksp}$, which are implicit when we write~\eqref{eq:phi0}) the action
of $\ee^{A+\tA}$ must generate a generic element of the orbit under
$\E7$.

Note that we could also define a real object $\lambda=\re\phi$
\begin{equation}
   \lambda = \ee^{A+\tA}
      (\ksp^\alpha\ksp^\beta,0,\bar{\ksp}_\alpha\bar{\ksp}_\beta,0) ,
\end{equation}
which also manifestly defines the same $\SU(7)$ structure. Let $N(E)$
be the $\rep{912}$ representation space based on the EGT $E$, at each
point $x\in M_7$, we can view $\lambda$ as an embedding of the coset
space
\begin{equation}
   \lambda : \frac{\E7}{\SU(7)}\times\bbR^+ \hookrightarrow N(E) ,
\end{equation}
where the $\bbR^+$ factor simply corresponds to a rescaling of
$\lambda$. We will call this orbit subspace $\Sigma$. There should
then be a natural complex structure on $\Sigma$ which allows one to
define the holomorphic $\phi$. This is in analogy to $\Upsilon^\pm$
and $\Phi^\pm$ in generalised geometry. Note that $\lambda$ far from
fills out the whole of the $\rep{912}$ representation space. Rather we
are considering a very particular orbit. One could always write down
the particular non-linear conditions which define the orbit, that is,
the analogues of~\eqref{eq:compat}.

Finally let us also consider how the supergravity fields decompose
under the $\SU(7)$ subgroup and how these correspond to different
$N=1$ multiplets. We have for $\SU(7)\subset\SU(8)$
\begin{equation}
\begin{aligned}
   \rep{8} &= \rep{7} + \rep{1} , &
   \rep{35} &= \rep{35} , \\
   \rep{28} &= \rep{21} + \rep{7} , &\qquad
   \rep{56} &= \rep{35} + \rep{21} .
\end{aligned}
\end{equation}
This means we can arrange the degrees of freedom as in
table~\ref{tab:SU7}.
\TABLE[h]{
   \begin{tabular}{lll}
      multiplet & $\SU(7)$ rep & fields \\
      \hline
      chiral & $\rep{35}$
         & $g_{mn}$, $A_{mnp}$, $\tA_{m_1\dots m_7}$, $\Psi_m$ \\
      vector & $\rep{21}$
         & $g_{\mu m}$, $A_{\mu np}$, $\Psi_m$ \\
      spin-$\tfrac{3}{2}$ & $\rep{7}$
         & $g_{\mu m}$, $A_{\mu np}$, $\Psi_\mu$ \\
      gravity & $\rep{1}$
         & $g_{\mu\nu}$, $\Psi_\mu$
   \end{tabular}
   \caption{Multiplet structure under $\SU(7)$}
   \label{tab:SU7}
}
Note that the coset space $\E7/\SU(7)$ actually decomposes into
$\rep{35}+\rep{7}+\bar{\rep{35}}+\bar{\rep{7}}$. Thus there are more
degrees of freedom in $\lambda$ than chiral degrees of freedom. The
same phenomenon appears in the type II case and is associated to the
gauge freedom of the extra spin-$\tfrac{3}{2}$ multiplets. One
solution is to assume in a given truncation of the theory that there
are no $\rep{7}$ degrees of freedom. Note that in this picture we
expect there to be a natural K\"ahler metric on the coset space
$\Sigma=\E7/\SU(7)\times\bbR^+$ corresponding to the K\"ahler
metric on the chiral scalar field space of $N=1$
theories~\cite{inprog}.


\section{Application: the effective superpotential}
\label{sec:W}


In the previous section we found the objects defining the $\SU(7)$
structure relevant to $N=1$ reformulations of eleven-dimensional
supergravity. The elements $\phi\in\Sigma$ should correspond to the
chiral multiplet degrees of freedom. As such there should be an
analogue of the four-dimensional superpotential $W$, as a holomorphic
function of $\phi$. In this section, we will derive the generic
structure of $W$ and show that it can be written in an $\E7$ covariant
form. This is the analogue of the corresponding generalised geometry
calculation in the case of type II given in~\cite{GLW1,GLW2}. Note
that the structure of $W$, for the special case of a $G_2$ structure
was previously derived in~\cite{HM}.


\subsection{Generic form of the effective superpotential}
\label{sec:genW}

We will read off $W$ from the variation of the four-dimensional
gravitino. Recall that the $N=1$ gravitino variations are given by
\begin{equation}
\label{eq:4dsusy}
   \delta \psi_{\mu +} = \nabla_{\mu} \theta_{+}
      + \tfrac{1}{2}\ii\ee^{K/2} W \gamma_{\mu} \theta_{-} + \dots ,
\end{equation}
where $W$ is the superpotential and $K$ the K\"ahler potential. The
expressions for $\ee^{K/2}W$ can then be derived directly from the
eleven-dimensional gravitino variation (see appendix~\ref{app:sugra}
for our conventions)
\begin{equation}
\label{susy11}
   \delta \Psi_M = \nabla_M \epsilon + \tfrac{1}{288}\left(
      \Gamma_M{}^{NPQR} - 8 \delta_M^N\Gamma^{PQR}\right)F_{MNPQ}
      + \dots ,
\end{equation}
where the dots denote terms depending on $\Psi_M$.

We must first identify the correctly normalised four-dimensional
gravitino $\psi_\mu$. A naive decomposition $\Psi_{M} = (\Psi_{\mu},\Psi_m)$ and
identifying $\psi_\mu$ as part of $\Psi_\mu$, leads to cross-terms in
the kinetic energy, so instead we first need to diagonalise the
four-dimensional gravitino kinetic energy term. This requires the
following shift
\begin{equation}
   \tilde{\Psi}_\mu := \Psi_\mu + \tfrac{1}{2}\Gamma_\mu\Gamma^m\Psi_m .
\end{equation}
One further has to rescale by a factor of $\ee^E$, and hence identify the
four-dimensional gravitino $\psi_\mu$ as the $\SU(7)$ singlet part
\begin{equation}
   \tilde{\Psi}_\mu = \ee^{E/2} \psi_{\mu +}\otimes{\ksp^c} +
       \ee^{E/2} \psi_{\mu +}^c \otimes\ksp + \dots
\end{equation}
where the dots denote non-singlet terms. This rescaling by $\ee^{E/2}$
is the reason for adopting the conventions in the spinor decomposition
given in~\eqref{eq:spinor-decomp}. Given we can rescale ${\ksp^c}$
by including factors in $\theta_+$, we can always choose a
normalisation
\begin{equation}
\label{norm}
   \bar\ksp\ksp = 1 .
\end{equation}
This allows us to introduce the projectors
\begin{equation}
\begin{aligned}
   \Pi_+ &:=  \tfrac{1}{2}(1+\gfive) \otimes {\ksp^c}\bar{\ksp}^c \\
   \Pi_- &:= \tfrac{1}{2}(1-\gfive) \otimes \ksp\bar\ksp
\end{aligned}
\end{equation}
such that
\begin{equation}
   \ee^{-E/2}\Pi_+ \tilde{\Psi}_\mu =  \psi_{\mu +} \otimes {\ksp^c} .
\end{equation}

It is now straightforward to calculate $\delta\psi_\mu$ in terms of
${\ksp^c}$, $\FF$ and $\tFF$. By definition
\begin{equation}
\begin{aligned}
   \delta\psi_\mu\otimes{\ksp^c} &= \ee^{-E/2}\Pi_+\delta\tilde{\Psi}_\mu
      = \ee^{-E/2}\Pi_+\left(\delta\Psi_\mu
         +\tfrac{1}{2}\Gamma_\mu\Gamma^m\delta\Psi_m\right) \\
      &= \nabla_{\mu} \theta_{+}\otimes{\ksp^c} + \tfrac{1}{2}\ii\ee^{K/2}W\gamma_\mu\theta_-\otimes{\ksp^c} ,
\end{aligned}
\end{equation}
which gives
\begin{equation}
   \ee^{K/2}W = \tfrac{1}{4}\ii\ee^E\left(
            4\bar{\ksp}^c\gamma^m\nabla_m\ksp
            + \tfrac{1}{4!}\FF_{mnpq}\bar{\ksp}^c\gamma^{mnpq}\ksp
            - \ii *_7\tFF \bar{\ksp}^c\ksp
            \right) .
\end{equation}
We have used the fact that $\bar{\ksp}^c\gamma^m\ksp=0$ identically to
remove $\nabla_mE$ terms. This expression can be put in a more
standard form by writing:
\begin{equation}
   \ksp\bar{\ksp}^c = \tfrac{1}{8} (\bar{\ksp}^c\ksp)\id
       - \tfrac{1}{8\cdot 3!}(\bar{\ksp}^c\gamma_{mnp}{\ksp^c})\gamma^{mnp} .
\end{equation}
Thus, since $\bar{\ksp}^c\gamma^m\ksp=\bar{\ksp}^c\gamma^{mn}\ksp=0$,
we have
\begin{equation}
\begin{split}
   (\bar{\ksp}^c\gamma^m\nabla_m\ksp)(\bar{\ksp}^c\ksp)
      &= \bar{\ksp}^c\gamma^m\nabla_m(\ksp\bar{\ksp}^c)\ksp \\
      &=  \tfrac{\ii}{8\cdot 3!}\nabla_m\varphi_{npq}
           (\bar{\ksp}^c\gamma^m\gamma^{npq}\ksp) \\
      &= - \tfrac{\ii}{8\cdot 4!}(\dd\varphi)_{mnpq}(*_7\varphi)^{mnpq} \\
      &= - \tfrac{\ii}{8}*_7(\varphi\wedge\dd\varphi) ,
\end{split}
\end{equation}
where we used~\eqref{eq:bilins}. Similarly,
\begin{equation}
   \tfrac{1}{4!}\FF_{mnpq}\bar{\ksp}^c\gamma^{mnpq}\ksp
      = - *_7 (\FF\wedge\varphi) .
\end{equation}
Hence
\begin{equation}
\label{eq:Wfinal}
\begin{split}
   *_7\, \ee^{K/2} W &= \frac{1}{8}\ee^{E}\left[
         \frac{1}{\bar{\ksp}^c\ksp}\varphi\wedge\dd\varphi
         - 2\ii\FF\wedge\varphi
         + 2\tFF \bar{\ksp}^c\ksp \right] \\
      &= \frac{1}{8}\bar{\ksp}^c\ksp\ee^{E}\left[
         \dd(\tphi-\ii A)\wedge(\tphi-\ii A)
         + 2\dd\tA - \ii\dd(A\wedge\tphi)\right]
\end{split}
\end{equation}
where we have introduced the renormalised
$\tphi=\varphi/\bar{\ksp}^c\ksp$.

In the case where ${\ksp^c}=\ksp$ we have a global $G_2$ structure, our
normalisation convention~\eqref{norm} implies that
$\bar{\ksp}^c\ksp=1$ and one finds that~\eqref{eq:Wfinal} agrees with
that derived in~\cite{HM}. The generic $\SU(7)$ case differs form the
simple $G_2$ case as through the pre-factor $\bar{\ksp}^c\ksp$ and the
fact that $\tilde\varphi$ is no longer real.



\subsection{An $\E7$ covariant expression}
\label{sec:covW}

In this section we show that one may rewrite the superpotential
term~\eqref{eq:Wfinal} in a manifestly $\E7$ invariant form using the
$\SU(7)$ structure $\phi\in\rep{912}$. This is the analogue of the
$O(6,6)$ invariant expressions for the $N=2$ prepotentials given
in~\cite{GLW2} for type IIA and IIB theories compactified on
$\SU(3)\times\SU(3)$ structure backgrounds.

We first need to introduce an embedding of the derivative operator
into an $\E7$ representation. Given the $\GL(7)$ decomposition of the
EGT given in~\eqref{eq:E0def}, we see that, assuming for the moment we
have a metric $g$, we can introduce an operator
\begin{equation}
   D = (D^{ab}, D_{ab}) \in \rep{56}
\end{equation}
with
\begin{equation}
\label{eq:Ddef}
   D^{mn} = D^{m8} = D_{mn} = 0 , \qquad
   D_{m8} = (\det g)^{1/4}\nabla_m .
\end{equation}
Transforming to spinor indices gives
\begin{equation}
   D = (D^{\alpha\beta},\bar{D}_{\alpha\beta})
      = (-\tfrac{1}{2\sqrt{2}}\gamma^{m\,\alpha\beta}\nabla_m,
          \tfrac{1}{2\sqrt{2}}\gamma^m_{\alpha\beta}\nabla_m) .
\end{equation}
Given the derivative operator we can define its action on
$\phi\in\rep{912}$. In particular using the product between $\rep{56}$
and $\rep{912}$ (see~\eqref{eq:Xphi} or~\eqref{eq:XphiSU8}) we can
define an object in the $\rep{133}$ representation which we denote as
$(D\cdot\phi)^{AB}=D_C\phi^{C(AB)}$ (where indices are raised and
lowered using the symplectic structure $\Omega_{AB}$).

The claim is that the superpotential can be written as
\begin{equation}
\label{eq:E7W}
   (D\cdot\phi)\,\phi
      = - \left(\frac{3}{4\sqrt{2}}\,\ee^{-E}\ee^{K/2}W\right) \phi ,
\end{equation}
where $(D\cdot\phi)\,\phi\in\rep{912}$ denotes the adjoint action of
$D\cdot\phi$ on $\phi$ itself. The statement is that
$(D\cdot\phi)\,\phi$ is itself an $\SU(7)$ singlet, proportional to
$\phi$ and $W$ is related to the constant of proportionality. Note
that we expect $W$ to be a holomorphic function of $\phi$, since
$\phi$ encodes the chiral multiplet scalar fields, whereas the
K\"ahler potential should be a function of both $\phi$ and its complex
conjugate. This suggests we should identify $\ee^E$ and $\ee^{K/2}$,
or equivalently $\ee^{-K}=\textrm{const.}\times\sqrt{\det g}$. We
return to this briefly in section~\ref{sec:concl}.

To show~\eqref{eq:E7W} requires two steps. First recall that we defined
$\phi=\ee^{A+\tA}\phi_0$. Writing $h=\ee^{A+\tA}\in\E7$, one can then
define a new operator $D_0$ with a connection taking values in the
adjoint of $\E7$ by conjugating by $h$. Making the $\E7$ indices
explicit, we define
\begin{equation}
   D_0^{AB}{}_C = D^A\delta^B{}_C + \kappa^{AB}{}_C ,
\end{equation}
where $\kappa^{AB}{}_C=(h^{-1})^B{}_ED^Ah^E{}_C$. One then has
$D\cdot\phi=\ee^{A+\tA}(D_0\cdot\phi_0)$ where
\begin{equation}
\begin{aligned}
   (D_0\cdot\phi_0)^{AB} &= \Omega_{CD} \Big(
            D^{C}\phi_0^{D(AB)} \\ &
        + \kappa^{CD}{}_E\phi_0^{E(AB)}
        + \kappa^{C(A}{}_E\phi_0^{DE|B)}
        + \kappa^{C(B}{}_E\phi_0^{D|A)E} \Big) ,
\end{aligned}
\end{equation}
where we have used the fact that $h^A{}_BD^B=D^A$. The
expression~\eqref{eq:E7W} can thus be rewritten as
\begin{equation}
   (D_0\cdot\phi_0)\,\phi_0
      = - \left(\frac{3}{4\sqrt{2}}\,\ee^{-E}\ee^{K/2}W\right) \phi_0 .
\end{equation}

It is then straightforward to calculate
$(D_0\cdot\phi_0)\,\phi_0$. Using the Hadamard formula
\begin{equation}
   \ee^P Q \ee^{-P} = Q + [P,Q] + \tfrac{1}{2}[P,[P,Q]] + \dots ,
\end{equation}
for operators $P$ and $Q$, one can calculate $D_0$. Given the commutator
algebra~\eqref{eq:A-algebra}, we have
\begin{equation}
\begin{split}
   \ee^{-A-\tA}\nabla_m\ee^{A+\tA}
      &= \nabla_m + \nabla_m(A+\tA)
         + \tfrac{1}{2}[\nabla_m(A+\tA),A+\tA] + \dots  \\
      &= \nabla_m + \nabla_mA+\nabla_m\tA
         -\tfrac{1}{2}\nabla_mA\wedge A \\
      &:= \nabla_m + \kappa_m .
\end{split}
\end{equation}
Note that this expression truncates at quadratic order. Again the
connection $\kappa_m$ takes values in the $\E7$ Lie algebra. In the
$\hg$ basis $\phi_0$ takes the form~\eqref{eq:phi0}. Hence
\begin{equation}
\begin{aligned}
   (\nabla_m+\kappa_m)\phi_0{}^{\alpha\beta}
      &= \nabla_m(\ksp^\alpha\ksp^\beta)
         + (\kappa_m{}^\alpha{}_\gamma\ksp^\gamma)\ksp^\beta
         + \ksp^\alpha(\kappa_m{}^\beta{}_\gamma\ksp^\gamma) , \\
   (\nabla_m+\kappa_m)\phi_0{}^{\alpha\beta\gamma}{}_\delta &= 0 , \\
   (\nabla_m+\kappa_m)\bar{\phi}_{0\;\alpha\beta} &= 0 , \\
   (\nabla_m+\kappa_m)\bar{\phi}_{0\;\alpha\beta\gamma}{}^\delta
      &= \kappa_{m\,\alpha\beta\gamma\epsilon}\ksp^\epsilon\ksp^\delta .
\end{aligned}
\end{equation}
so that
\begin{equation}
\begin{aligned}
   (D_0\cdot\phi_0)^\alpha{}_\beta &= \frac{3\ii}{8\sqrt{2}} \Big[
          \nabla_m(\ksp^\alpha\ksp^\gamma)\gamma^m{}_{\gamma\beta}
          \\ & \qquad \qquad
          + \kappa_m{}^\alpha{}_\gamma\ksp^\gamma
             \ksp^\delta\gamma^m{}_{\delta\beta}
          - \ksp^\alpha\gamma^m{}_{\beta\gamma}
             \kappa_m{}^\gamma{}_\delta\ksp^\delta
          - \ksp^\alpha\gamma^{m\,\gamma\delta}
             \kappa_{m\,\gamma\delta\beta\epsilon}\ksp^\epsilon
             \Big] \\
   (D_0\cdot\phi_0)_{\alpha\beta\gamma\delta}
      &= -\frac{\ii}{2\sqrt{2}} \kappa_{m\,[\alpha\beta\gamma|\epsilon}
         \ksp^\epsilon\gamma^m{}_{\delta]\theta}\ksp^\theta .
\end{aligned}
\end{equation}
Finally, again using the identity
$\ksp^\alpha\gamma^m_{\alpha\beta}\ksp^\beta=0$ we find
\begin{equation}
\label{eq:nearly}
\begin{aligned}
   \left[(D_0\cdot\phi_0)\,\phi_0\right]^{\alpha\beta}
      &= -\frac{3\ii}{4\sqrt{2}} \left(
         \ksp^\gamma\gamma^m{}_{\gamma\delta}\nabla_m\ksp^\delta
         + \ksp^\gamma\gamma^m_{\gamma\delta}
            \kappa_m{}^\delta{}_\epsilon\ksp^\epsilon
         \right)\ksp^\alpha\ksp^\beta \\
      &= -\frac{3\ii}{4\sqrt{2}}\Big[
         \bar{\ksp}^c\gamma^m(\nabla_m+\kappa_m)\ksp
         \Big]\phi_0^{\alpha\beta} ,
\end{aligned}
\end{equation}
with all other components vanishing. We see that, as claimed,
$(D_0\cdot\phi_0)\,\phi_0$ is proportional to $\phi_0$.

Finally we recall that $\kappa_m$ corresponded to an ``$A$-shift'' of
$\nabla_mA$ and a ``$\tA$-shift'' of
$\nabla_m\tA-\frac{1}{2}\nabla_mA\wedge A$. Only the $\hg$ adjoint
component $\kappa_m{}^\alpha{}_\beta$ survives
in~\eqref{eq:nearly}. Using the decomposition~\eqref{eq:SU8decomp}
together with the definitions~\eqref{eq:AA'def} we find this component
is given by
\begin{equation}
   \kappa_m = \tfrac{1}{4\cdot 4!}(\nabla_m A)_{npq} \gamma^{npq}
      - \tfrac{1}{4}\ii\big[ *_7\big( \nabla_m\tA
         -\tfrac{1}{2}\nabla_mA\wedge A \big)\big]_n\gamma^n
\end{equation}
and hence
\begin{equation}
\begin{aligned}
   \bar{\ksp}^c\gamma^m(\nabla_m + \kappa_m)\ksp
      &=  \bar{\ksp}^c\gamma^m\nabla_m\ksp
            + \tfrac{1}{4\cdot 4!}\FF_{mnpq}\bar{\ksp}^c \gamma^{mnpq}\ksp
            - \ii(*_7\tFF) \bar{\ksp}^c\ksp \\
      &= -\ii\ee^{-E}\ee^{K/2}W ,
\end{aligned}
\end{equation}
as required.

We now return briefly to a subtlety in the definition of $D$. As
written~\eqref{eq:Ddef}, $D$ is only defined given a metric
$g$. However $\phi$ defines an $\SU(7)$ structure on $E$ and hence a
metric $g$ (and form fields $A$ and $\tA$). Thus, as written, we can
think of $D$ as defined in terms of $\phi$. This is in contrast with
the type II $\stt$ case. There the exterior derivative defined a
natural generalised Dirac operator $\dd:S^\pm(E)\to S^\mp(E)$
independent of the structure (or specifically any metric). It is also
in contrast with the final result: the final superpotential can be
written~\eqref{eq:Wfinal} using only the exterior derivative.

The indication is that one can actually define $D$ independently of
the metric, such that it has a sensible action on $\phi$. In this
sense the differential EGG is set up before introducing any
structure, as is that case in generalised geometry. One would also
expect that such a $D$ is dual to the ECB in the same way that the
exterior derivative on $S^\pm(E)$ is dual to the ordinary Courant
bracket~\cite{H-brack}. A subtlety in the generalised geometrical case
is that the isomorphism between $S^\pm(E)$ and $\Lambda^*T^*M$ is
not unique: the natural isomorphism is to
$(\Lambda^dT^*M)^{-1/2}\otimes\Lambda^*T^*M$~\cite{Gualtieri}. Thus to
define the exterior derivative one must rescale by something in
$(\Lambda^dT^*M)^{-1/2}$. We expect something similar in the
definitions of $\phi$ and $D$, removing the need for $(\det g)^{1/4}$
in~\eqref{eq:Ddef}.


\section{Conclusions}
\label{sec:concl}


In this paper we have discussed an extension of generalised geometry
applicable to eleven-dimensional supergravity and for which the
symmetry group is the continuous U-duality group $E_{d(d)}$. The
general form of such constructions was recently discussed by
Hull~\cite{chris}. Here we specifically focused on generic $N=1$ flux
compactifications to four dimensions, for which the relevant symmetry
group is $\E7$. We showed that $N=1$ supersymmetry implies that there
is $\SU(7)$ structure on this ``exceptional generalised  geometry''
(EGG) defined by an element  $\phi$ in a particular orbit in the
$\rep{912}$ representation of $\E7$. This is the analogue of the pair
of generalised spinors $\Phi^\pm$, each defining a generalised complex
structure, which characterise $N=2$ type II backgrounds. In the
four-dimensional theory it encodes the chiral multiplet scalar degrees
of freedom. As an application we showed that the superpotential for
generic $N=1$ flux compactifications could be written as an
$\E7$-invariant homogeneous, holomorphic function of $\phi$.

\TABLE[b]{
\begin{tabular}{ccc|ccc}
   \multicolumn{2}{c}{generalised geometry}
      &&&
      \multicolumn{2}{c}{exceptional generalised geometry} \\
   \multicolumn{2}{c}{\footnotesize $E_0=TM\oplus TM^*$}
      &&&
      \multicolumn{2}{l}{\footnotesize $E_0=TM \oplus \Lambda^2T^*M \oplus
        \Lambda^5T^*M  \oplus (T^*M \otimes\Lambda^7T^*M)$}  \\*[5pt]
   structure & group &&& structure & group \\*[2pt]
   $\met$ & $O(d,d)$ &&&
   $(\Omega,q)$ & $\E7$ \\
   $\Pi$& $\GL(d)\times \GL(d)$ &&&
   $J$ & $\GL(28,\bbC)$ \\
   $G$ & $O(2d)$ &&&
   $G$ & $O(56)$ \\
   $(\met,\Pi)$ or $(\met,G)$ & $O(d)\times O(d)$ &&&
   $(\Omega,q,J)$ or $(\Omega,q,G)$ & $\hg$ \\
\end{tabular}
\caption{Comparing generalised and exceptional generalised geometry}
\label{tab:compare}
}
In fact, almost all the objects appearing in generalised geometry have
analogues in EGG. There is an exceptional generalised tangent space
$E$, now combining vectors, two-forms, five-forms and an eight-index
tensor. This is twisted by gerbes which capture the topological
information encoded in the patching of the three-form supergravity
potential $A$ and its dual $\tA$ -- the analogues of the NS--NS
$B$-field. There is also a natural exceptional Courant bracket,
encoding the differential structure on $E$. The seven-dimensional
metric $g$ and potentials $A$ and $\tA$ then combine to define an
$\hg$ structure on $E$. This is the analogue of the generalised
metric, combining metric and $B$-field, that defines an $O(d)\times
O(d)$ structure in generalised geometry. If the background has $N=1$
supersymmetry this structure is further refined to $\SU(7)$, while
for type II six-dimensional $N=2$ backgrounds the generalised
structure is $\stt\subset O(6)\times O(6)$. These parallels are
partially summarised in table~\ref{tab:compare}.

There are a number of obvious extensions of this work one would like
to consider. Directly related to the results described here are the
questions, first, of the definition of the derivative operator $D$ and,
secondly, of the form of the K\"ahler potential~\cite{inprog}. As
discussed in section~\ref{sec:covW}, we expect that the
definition~\eqref{eq:Ddef} of $D$ can be replaced with one written in
terms of the ordinary partial derivative without need for a
metric. This should, in an appropriate sense following~\cite{H-brack},
be the dual of the exceptional Courant bracket defined
in~\eqref{eq:ECB}.

As for the superpotential the K\"ahler potential (or rather the
K\"ahler metric) can be calculated directly by identifying the
four-dimensional kinetic terms in the decomposition of the
eleven-dimensional theory. One would again expect the potential to be
a $\E7$ invariant. We know from the work on type II
theories~\cite{GLW1,GLW2} that $\ee^{-K}$ is proportional to the
metric density $\sqrt{\det g}$. A similarly relation here would be
compatible with the expression~\eqref{eq:E7W} for $W$ being a
holomorphic function of $\phi$.

In the type II case we also find~\cite{GLW1,GLW2} that $\ee^{-K}$ is
proportional to the Hitchin functional~\cite{GCY,Hfunc}. This is true
both for the ordinary geometrical case for backgrounds parametrised by
an $\SU(3)$ structure and for the extension to the generalised geometrical
Hitchin functional for $\SU(3,3)$ structures $\Phi^\pm$. Hitchin has
already introduced a functional for a conventional $G_2$
structure. One expects that the $\E7$-invariant K\"ahler potential
should be the EGG generalisation of this functional for $\SU(7)$
structures defined by $\phi$ which includes the potential $A$ and
$\tA$ degrees of freedom. This also should be the natural
generalisation of the action for topological M-theory~\cite{topM}. As
the generalised geometrical functionals necessarily arose at one-loop
in the topological B-model~\cite{PW}, one conjecture is that the
putative EGG $\SU(7)$ functional would appear at one-loop in
topological M-theory. However, note that the extension could also be
to the generalised geometrical $G_2\times G_2$ functional~\cite{BMSS}.

An obvious question to address is to consider not the four-dimensional
effective theory but the on-shell supersymmetric
backgrounds~\cite{inprog}. For type II theories, satisfying the
six-dimensional Killing spinor equations is equivalent to simple
differential conditions on the $\stt$ structures
$(\Phi^+,\Phi^-)$~\cite{GMPT,JW,GMPT2,JW2}. One expects a very
similar relation in EGG for the $\SU(7)$ structure $\phi$ using the
operator $D$. These ``integrability'' conditions on the generalised
geometry and EGG structures are the generalisations of the
$G$-structure and intrinsic torsion classification of supersymmetric
backgrounds~\cite{G-struc}.

One can also repeat the analysis here for compactifications of
eleven-dimensional supergravity to other dimensions, or for that
matter for type II theories where the EGG ``geometrises'' the
Ramond--Ramond degrees of freedom~\cite{chris}. There will also be
relations between backgrounds in different dimensions. For instance
the four-dimensional type II $N=2$ backgrounds~\cite{GLW1,GLW2} should
be encoded in the dimensional reduction of the $N=1$ backgrounds
discussed here. In particular, one notes that in terms of the
generalised structures $O(6,6)\subset\E7$ and for supersymmetry
$\stt\subset\SU(7)$.

Let us end by noting two further connections. Hull's work~\cite{chris}
was partly motivated by the existence of non-geometrical
backgrounds~\cite{Tfold}--\cite{Hull:2007jy}. The exceptional generalised tangent
space~\eqref{eq:twistE} is patched not by generic elements of $\E7$
but only elements in the subgroup $\GL(7)\ltimes\SG$, that is the
usual geometrical patching of the tangent spaces together with $A$-
and $\tA$-shifts (by exact forms). It is very natural to extend the
twisting to generic $\E7$ bundles as discussed in~\cite{chris}. Such
spaces cannot directly describe non-geometrical backgrounds since the
underlying space $M$ is still a conventional manifold. Nonetheless
they are closely connected to the doubled T-fold and U-fold geometries
of~\cite{Tfold}. It was argued recently~\cite{DPST} that the generic
four-dimensional $N=8$ gauged supergravity theories arise from
compactification on a 56-dimensional ``megatorus'' U-fold. Interestingly
such generic supergravities are encoded by an embedding
tensor~\cite{dWST} which, like $\phi$, lies in the $\rep{912}$
representation of $\E7$. Here, we have focused on $N=1$
theories rather than $N=8$ in four dimensions. The appearance of the
$\rep{912}$ representation is nonetheless indirectly connected to the
embedding tensor. For $N=8$, the embedding tensor appears (via the
$T$-tensor) in the supersymmetry variations of the eight gravitinos
and 56 spin-$\frac{1}{2}$ fields. Decomposing under $\SU(8)$ the
$\rep{36}$ representation appears in the former and the $\rep{420}$ in
the latter. In order to define an $N=1$ theory, we further decompose under
$\SU(7)$. The $\rep{36}$ representation decomposes as
$\rep{36}=\rep{1}+\rep{7}+\rep{28}$. The first term goes with the
$N=1$ gravitino and corresponds to the superpotential
term~\eqref{eq:4dsusy}. In addition the $\rep{420}$ representation
decomposes as $\rep{420}=\rep{224}+\rep{140}+\rep{35}+\rep{21}$. The
chiral multiplets are the $\rep{35}$ representation (see
table~\ref{tab:SU7}), and so the $\rep{35}$ term in the $T$-tensor
corresponds the derivative of the superpotential with respect to the
chiral scalars. Thus the structure $\phi$ and the $N=8$ embedding
tensor share a common $\rep{35}$ representation under the $\SU(7)$
decomposition.


\acknowledgments

We would like to thank Jerome Gauntlett, Jan Louis and, in particular,
Mariana Gra\~{n}a and Chris Hull for helpful discussions.
P.~P.~P. thanks the FCT (part of the Portuguese Ministry of Education)
for financial support under scholarship SFRH/BD/10889/2002. D.~W. is
supported by a Royal Society University Research Fellowship and thanks
the Aspen Center for Physics, CEA/Saclay and the Isaac Newton
Institute for Mathematical Sciences for hospitality during the
completion of this work.


\appendix


\section{Conventions}
\label{app:convs}


\subsection{Eleven-dimensional supergravity}
\label{app:sugra}

We adopt conventions where the eleven-dimensional supergravity action
takes the form (see for instance~\cite{HM})
\begin{equation}
\label{action11}
   S = \frac{1}{2\kappa^2}\int_{M_{11}}
      \sqrt{-g}\left(R - \bar{\Psi}_M\Gamma^{MNP}D_N\Psi_P\right)
      - \tfrac{1}{2} F \wedge *F
      - \tfrac{1}{6} A \wedge F \wedge F + \dots ,
\end{equation}
while the variation of the gravitino $\Psi_M$ is given by
\begin{equation}
   \delta \Psi_M = \nabla_M \epsilon + \tfrac{1}{288}\left(
      \Gamma_M{}^{NPQR} - 8 \delta_M^N\Gamma^{PQR}\right)F_{MNPQ}
      + \dots .
\end{equation}
(The dots represent four-fermi terms and terms coupling $\Psi_M$ and
$F$.) Here $M,N=0,1,\dots,10$ are eleven-dimensional indices, the
metric $g$ has signature $(-,+,\dots,+)$ and $\Gamma_M$ are the
eleven-dimensional gamma matrices satisfying
\begin{equation}
   \left\{ \Gamma_M, \Gamma_N \right\} = 2 g_{MN} \id ,
\end{equation}
with $\Gamma_{M_1\dots M_{11}}=\id\epsilon_{M_1\dots M_{11}}$ where
the volume form $\epsilon$ is given by $\epsilon=\sqrt{-g}\,\dd
x^0\wedge\dots\wedge\dd x^{10}$. The spinors $\epsilon$ and $\Psi_M$
are Majorana. Given the intertwining relation
$-\Gamma_M^T=C^{-1}\Gamma_MC$, we define the conjugate
spinor  $\bar{\Psi}_M=\Psi_M^TC^{-1}$. Note that the equation of
motion and Bianchi identity for the flux $F$ read
\begin{equation}
\label{eq:Feom}
   \dd * F + \tfrac{1}{2}F\wedge F = 0 , \qquad
   \dd F = 0 .
\end{equation}
%


\subsection{$\Cliff(7,0)$ and seven-dimensional spinors}
\label{app:7d}

Let us also fix our conventions for $\Spin(7)$ (see also Appendix C
of~\cite{CJ}). The Clifford algebra $\Cliff(7,0;\bbR)$ is generated by
the gamma matrices $\gamma_m$ with $m=1,\dots,7$ satisfying
\begin{equation}
      \{ \gamma_m, \gamma_n \} = 2 g_{mn}\id .
\end{equation}
One finds $\Cliff(7,0;\bbR)\simeq\GL(8,\bbC)$ and hence the spinor
representation of the Clifford algebra is complex and
eight-dimensional. We define the intertwiners $A$ and $C$ by
\begin{equation}
   \gamma_M^\dag = A\gamma_m A^{-1} , \qquad
   -\gamma_m^T = C^{-1}\gamma_m C ,
\end{equation}
with $A^\dag=A$, $C^T=C$ and such that $-\gamma_m^*=D^{-1}\gamma_mD$
with $D=CA^T$. Given a spinor $\ksp$ we define the conjugate spinors
\begin{equation}
   \bar{\ksp}^c={\ksp^c}^\dag A , \qquad
   \ksp = D {\ksp^c}^* .
\end{equation}
Furthermore, we write
\begin{equation}
   \gamma_{m_1\dots m_7}=\gamma_{(7)}\epsilon_{m_1\dots m_7}
\end{equation}
with $\epsilon=\sqrt{g}\dd x^1\wedge \dots \wedge\dd x^7$. Note that
one can choose the gamma matrices such that $\gamma_{(7)}=\ii$. The
intertwiner $A$ provides an hermitian metric on the spinor space,
which is invariant under the subgroup $\SU(8)\subset\Cliff(7,0;\bbR)$,
with a Lie algebra spanned by
$\{\gamma_{m_1,m_2},\gamma_{m_1m_2m_3},\gamma_{m_1\dots
  m_6},\gamma_{m_1\dots m_7}\}$.

The even part of the Clifford algebra generated by the $\gamma_{mn}$
has $\Cliff(7,0;\bbR)_{\text{even}}\simeq\GL(8,\bbR)$ and hence a real
spinor representation with $\ksp=\ksp^c$. Thus the spin group
$\Spin(7)\subset\Cliff(7,0;\bbR)_{\text{even}}$ similarly has a real
spinor representation. For real spinors, $\bar{\ksp}={\ksp}^TC^{-1}$,
and $C^{-1}$ provides metric on the spin space. This is invariant
under a $\Spin(8)$ group with Lie algebra spanned by
$\{\gamma_{m_1m_2},\gamma_{m_1\dots m_6}\}$. This can alternatively be
described by, for $a,b=1,\dots 8$
\begin{equation}
\label{eq:gamma8def}
   \tgam_{ab}
      = \begin{cases}
         \,\,\,\,(\det g)^{-1/4}\gamma_{mn} & \text{if $a=m$, $b=n$} \\
          \,\,\,\,(\det g)^{1/4}\gamma_m\gamma_{(7)} & \text{if $a=m$, $b=8$} \\
         -(\det g)^{1/4} \gamma_n\gamma_{(7)} & \text{if $a=8$, $b=n$}
      \end{cases}
\end{equation}
which generate the $\Spin(8)$ Lie algebra with metric
\begin{equation}
   \tg_{ab} = (\det g)^{-1/4} \begin{pmatrix}
         g_{mn} & 0 \\
         0 & \det g
      \end{pmatrix} .
\end{equation}
Here we have introduced some factors of $\det g$ to match the form of
$\tg$ used elsewhere in the paper (in particular the decomposition
under $\GL(7,\bbR)$ given in section~\ref{app:GL7}). With these conventions,
the spinors $\ksp$ are of positive chirality with respect to $\Spin(8)$.

If we make the spinor indices explicit writing $\ksp^\alpha$ with
$\alpha=1,\dots,8$ we can raise and lower spinor indices using the
metric $C^{-1}$ so, for instance,
\begin{equation}
\label{eq:bigammas}
   \tgam_{ab\,\alpha\beta}
      = C^{-1}_{\alpha\gamma}\tgam_{ab}{}^\gamma{}_\beta  , \qquad
   \tgam_{ab}{}^{\alpha\beta}
      = \tgam_{mn}{}^\alpha{}_\gamma C^{\gamma\beta} .
\end{equation}
One also has the useful completeness relations, reflecting $\Spin(8)$
triality,
\begin{equation}
\label{eq:complete}
\begin{aligned}
   \tgam_{ab}{}^{\alpha\beta}\tgam^{ab}{}_{\gamma\delta}
      &= 16 \delta^{[\alpha}_{[\gamma}\delta^{\beta]}_{\delta]} , \\
   \tgam_{ab}{}^{\alpha\beta}\tgam^{cd}{}_{\alpha\beta}
      &= 16 \delta^{[c}_{[a}\delta^{d]}_{b]} .
\end{aligned}
\end{equation}


\section{The exceptional Lie group $\E7$}
\label{app:E7}


In this appendix we review some of the properties of the group $\E7$
relevant to this paper. A detailed definition of $\E7$ and an fairly
exhaustive description of its properties can be found in Appendix B
of~\cite{CJ} which itself refers to the original work by
Cartan~\cite{Cartan}.


\subsection{Definition and the $\rep{56}$ representation}
\label{app:E7def}

The group $\E7$ can be defined by its action on the basic
56-dimensional representation as follows. Let $W$ be a real
56-dimensional vector space with a symplectic product $\Omega$, then
$\E7$ is a subgroup of $\Symp(56,\bbR)$ leaving invariant a
particular quartic invariant $q$.

Explicitly one can define $\Omega$ and $q$ using the
$\SL(8,\bbR)\subset\E7$ subgroup. If $V$ is an eight-dimensional
vector space, on which $\SL(8,\bbR)$ acts in the fundamental
representation, then the $\rep{56}$ representation decomposes as
\begin{equation}
   \rep{56} = \rep{28} + \rep{28'}
\end{equation}
the $\rep{28}$ representation corresponding to $\Lambda^2V$ and the
$\rep{28'}$ to $\Lambda^2V^*$. Note that using
$\epsilon\in\Lambda^8V$, the totally antisymmetric form preserved by
the $\SL(8,\bbR)$ action on $V$, one can identify $\Lambda^2V^*$ with
$\Lambda^6V$. In summary, one identifies
\begin{equation}
\label{eq:Wdef}
   W = \Lambda^2V \oplus \Lambda^2V^*
\end{equation}
and writes $X\in W$ as the pair $(x^{ab},x'_{ab})$ where $a,b=1,\dots,8$.

The symplectic product $\Omega$ is then given by, where $A=1,\dots,56$
\begin{equation}
\label{eq:Odef}
   \Omega(X,Y) = \Omega_{AB} X^A Y^B
      = x^{ab}y'_{ab} - x'_{ab}y^{ab} ,
\end{equation}
and the quartic invariant $q$ is
\begin{equation}
\label{eq:qdef}
\begin{aligned}
   q(X) &= q_{ABCD} X^A X^B X^C X^D  \\
      &= x^{ab} x'_{bc} x^{cd} x'_{da}
         - \tfrac{1}{4} x^{ab}x'_{ab}x^{cd} x'_{cd} \\ & \qquad
         + \tfrac{1}{96} \left(
            \epsilon_{abcdefgh} x^{ab} x^{cd} x^{ef} x^{gh}
            + \epsilon^{abcdefgh} x'_{ab} x'_{cd}x'_{ef}x'_{gh} \right)
\end{aligned}
\end{equation}

In what follows it will be useful to use a matrix notation where we write
\begin{equation}
   X^A = \begin{pmatrix} x^{aa'} \\ x'_{aa'} \end{pmatrix}
\end{equation}
such that the symplectic form
\begin{equation}
   \Omega_{AB} = \begin{pmatrix}
         0 & \delta^b_a\delta^{b'}_{a'} \\
         \delta^a_b\delta^{a'}_{b'} & 0
      \end{pmatrix}
\end{equation}
Throughout it is assumed that \emph{all pairs of primed and unprimed
  indices $(a,a')$ etc are antisymmetrised}.


\subsection{The $\rep{133}$ and $\rep{912}$ representations}
\label{app:reps}

There are two other representations of interest in this paper. First is the
adjoint. By definition it is a 133-dimensional subspace $A$ of the Lie
algebra $\symp(56,\bbR)$. It decomposes under $\SL(8,\bbR)$ as
\begin{equation}
\begin{aligned}
   A &= (V\otimes V^*)_0 \oplus \Lambda^4V^* \\
   \mu &= (\mu^a{}_b, \mu_{abcd}) \\
   \rep{133} &= \rep{63} + \rep{70} ,
\end{aligned}
\end{equation}
where $(V\otimes V^*)_0$ denotes traceless matrices, so
$\mu^a{}_a=0$. The action on the $\rep{56}$ representation is given by
\begin{equation}
\begin{aligned}
   \delta x^{ab} &=
     \mu^a{}_cx^{cb} + \mu^b{}_cx^{ac} + \smu^{abcd} x'_{cd} , \\
   \delta x'_{ab} &=
     - \mu^c{}_ax'_{cb} - \mu^c{}_bx'_{ac} + \mu_{abcd} x^{cd} ,
\end{aligned}
\end{equation}
with $\smu^{a_1\dots a_4}=\frac{1}{4!}\epsilon^{a_1\dots
  a_8}\mu_{a_5\dots a_8}$. In terms of the matrix notation we have
$\delta X^A=\mu^A{}_BX^B$ with
\begin{equation}
   \mu^A{}_B = \begin{pmatrix}
            2\mu^a{}_b\delta^{a'}_{b'} & \smu^{aa'bb'} \\
            \mu_{aa'bb'} & - 2\mu^b{}_a\delta^{b'}_{a'}
         \end{pmatrix} .
\end{equation}
Note that $\mu^{AB}=\mu^A{}_C\Omega^{-1\, CB}$ is a symmetric
matrix. Taking commutators of the adjoint action gives the Lie algebra
$\mu^{\prime\prime}=[\mu,\mu']$
\begin{equation}
\begin{aligned}
   \mu^{\prime\prime\, a}{}_b &=
      ( \mu^a{}_c\mu^{\prime\, c}{}_b - \mu^{\prime\, a}{}_c\mu^c{}_b )
        + \tfrac{1}{3} ( \smu^{ac_1c_2c_3}\mu'_{bc_1c_2c_3}
           - \smu^{\prime\, ac_1c_2c_3}\mu_{bc_1c_2c_3} ) , \\
   \mu^{\prime\prime}_{abcd} &= 4 ( \mu^e{}_{[a}\mu'_{bcd]e}
        + \mu^{\prime\, e}{}_{[a}\mu_{bcd]e} ) .
\end{aligned}
\end{equation}

The other representation of interest in this paper is the
$\rep{912}$. The representation space $N$ decomposes under
$\SL(8,\bbR)$ as
\begin{equation}
\begin{aligned}
   N &= S^2V \oplus (\Lambda^3V\otimes V^*)_0
        \oplus S^2V^* \oplus (\Lambda^3V^*\otimes V)_0 \\
   \phi &= (\phi^{ab},\phi^{abc}{}_d,\phi'_{ab}, \phi'_{abc}{}^d) \\
   \rep{912} &= \rep{36} + \rep{420} + \rep{36'} + \rep{420'} ,
\end{aligned}
\end{equation}
where $S^nV$ denotes the symmetric product and $(\Lambda^3V\otimes
V^*)_0$ denotes traceless tensors, so $\phi^{abc}{}_c=0$. The adjoint
action of $\E7$ on $\phi$ is given by
\begin{equation}
\label{eq:muphi}
\begin{aligned}
   \delta\phi^{ab} &= \mu^a{}_c\phi^{cb} + \mu^b{}_c\phi^{ac}
      - \tfrac{1}{3}(
         \smu^{acde}\phi'_{cde}{}^b + \smu^{bcde}\phi'_{cde}{}^a ) , \\
   \delta\phi^{abc}{}_d &= 3\mu^{[a}{}_e\phi^{bc]e}{}_d
      - \mu^e{}_d\phi^{abc}{}_e
      + \smu^{abce}\phi'_{ed}
      + \smu^{ef[ab}\phi'_{efd}{}^{c]}
      - \smu^{efg[a}\phi'_{efg}{}^{b}\delta^{c]}_d , \\
   \delta\phi'_{ab} &= - \mu^c{}_a\phi'_{cb} - \mu^c{}_b\phi'_{ac}
      - \tfrac{1}{3}(\mu_{acde}\phi^{cde}{}_b
         + \mu_{bcde}\phi^{cde}{}_a ) , \\
   \delta\phi'_{abc}{}^d &= - 3\mu^e{}_{[a}\phi'_{bc]e}{}^d
      + \mu^d{}_e\phi'_{abc}{}^e
      + \mu_{abce}\phi^{ed}
      + \mu_{ef[ab}\phi^{efd}{}_{c]}
      - \mu_{efg[a}\phi^{efg}{}_{b}\delta_{c]}^d .
\end{aligned}
\end{equation}
In terms of $\Symp(56,\bbR)$ indices, we have $\phi^{ABC}$,
corresponding to the Young tableau {\scriptsize $\young(AC,B)$}
with $\phi^{ABC}\Omega_{AB}=0$. The different components are given by
\begin{equation}
\begin{aligned}
   \phi^{aa'bb'cc'} &= - \tfrac{1}{12}(
      \epsilon^{abb'cc'efg}\phi'_{efg}{}^{a'}
      - \epsilon^{baa'cc'efg}\phi'_{efg}{}^{b'} ) , \\
   \phi^{aa'bb'}{}_{cc'} &= 2\phi^{ab}\delta_c^{a'}\delta_{c'}^{b'}
      - \phi^{aa'b}{}_c\delta^{b'}_{c'}
      + \phi^{bb'a}{}_c\delta^{a'}_{c'} , \\
   \phi^{aa'}{}_{bb'}{}^{cc'} &= \phi^{ac}\delta^{a'}_b\delta^{c'}_{b'}
      - 2\phi^{aa'c}{}_b\delta_{b'}^{c'}
      - \phi^{cc'a}{}_b\delta^{a'}_{b'} ,
\end{aligned}
\end{equation}
with $\phi^{aa'}{}_{bb'}{}^{cc'} = - \phi_{bb'}{}^{aa'cc'}$ and
identical expressions for $\phi_{aa'bb'cc'}$ etc. but with raised and
lowered indices reversed.

Finally we will also need the tensor product
\begin{equation}
   \rep{56} \times \rep{912} = \rep{133} + \dots .
\end{equation}
In terms of $\Symp(56,\bbR)$ indices we have
$\mu^{AB}=X^C\Omega_{CD}\phi^{D(AB)}$, while in terms of $\SL(8,\bbR)$
components one finds
\begin{equation}
\label{eq:Xphi}
\begin{aligned}
   \mu^a{}_b &= \tfrac{3}{4}\left(
         x^{ac}\phi'_{cb} - x'_{bc}\phi^{ca}\right)
      + \tfrac{3}{4} \left(
         x^{cd}\phi'_{cdb}{}^a - x'_{cd}\phi^{cda}{}_b \right) , \\
   \mu_{abcd} &= - 3 \left(\phi'_{[abc}{}^ex'_{d]e}
         + \tfrac{1}{4!}\epsilon_{abcdm_1\dots m_4}
            \phi^{m_1m_2m_3}{}_ex^{m_4e} \right) .
\end{aligned}
\end{equation}
%


\subsection{A $\GL(7)$ subgroup}
\label{app:GL7}

As described in section~\ref{sec:EGT}, the tangent space structure
group embeds in the action of $\E7$ on the EGT. To make this embedding
explicit we must identify a particular
$\GL(7,\bbR)\subset\SL(8,\bbR)\subset\E7$ subgroup. In this appendix
we identify this group and give explicit expressions for part of the
$\E7$ action in terms of $\GL(7,\bbR)$ (that is spacetime tensor)
representations.

We start with the embedding of $\GL(7,\bbR)$ in $\SL(8,\bbR)$ given
by the matrix
\begin{equation}
   \begin{pmatrix}
         (\det M)^{-1/4} M^m{}_n & 0 \\ 0 & (\det M)^{3/4}
      \end{pmatrix} \in \SL(8,\bbR),
\end{equation}
where $M\in\GL(7,\bbR)$. If $\GL(7,\bbR)$ acts linearly on the
seven-dimensional vector space $F$ this corresponds to the
decomposition of the eight-dimensional representation space $V$ as
\begin{equation}
   V = (\Lambda^7F)^{-1/4}F \oplus (\Lambda^7F)^{3/4} .
\end{equation}
The $\rep{56}$ representation~\eqref{eq:Wdef} of $\E7$ then decomposes as\footnote{Note that, up to the $(\Lambda^7F^*)^{-1/2}$ factor, the same embedding of $\GL(7,\bbR)$ was identified in~\cite{West-decomp}. This followed from an analysis of $E_{11}$ representations in~\cite{West2}.}
\begin{equation}
   W =  (\Lambda^7F^*)^{-1/2} \otimes \left[
         F \oplus \Lambda^2F^* \oplus \Lambda^5F^*
         \oplus F^*\otimes(\Lambda^7F^*) \right] .
\end{equation}
We can write an element of $W$ as
\begin{equation}
   X = x + \omega + \sigma + \tau ,
\end{equation}
where $x\in (\Lambda^7F^*)^{-1/2}\otimes F$ etc. If we write the index
$m=1,\dots,7$ for the fundamental $\GL(7,\bbR)$ representation,  note
that, ignoring the tensor density factor $(\Lambda^7F^*)^{-1/2}$,
$\tau$ has the index structure $\tau_{m,n_1\dots n_7}$, where $n$
labels the $F^*$ factor and $n_1\dots n_7$ the $\Lambda^7F^*$
factor. We can make the identification between $\GL(7,\bbR)$ indices
and $\SL(8,\bbR)$ indices explicit by writing (again ignoring the
$(\Lambda^7F^*)^{-1/2}$ factor)
\begin{equation}
\begin{aligned}
   x^{m8} &= x^m & x^{mn} = \sigma^{mn}_{1\dots 7} \\
   x'_{m8} &= \tau_{m,1\dots 7} & x'_{mn} = \omega_{mn} ,
\end{aligned}
\end{equation}
where $\sigma^{mn}_{p_1\dots p_7}=
(7!/5!)\delta^m_{[p_1}\delta^n_{p_2}\sigma_{p_3\dots p_7]}$.

We can similarly decompose the $\rep{133}$ representation. We find
\begin{equation}
\begin{aligned}
   A &= (V\otimes V^*)_0 \oplus \Lambda^4V^* \\
     &= F\otimes F^* \oplus \Lambda^6F \oplus \Lambda^6F^*
        \oplus \Lambda^3F \oplus \Lambda^3F^* .
\end{aligned}
\end{equation}
We will be particularly interested in the action of the $\Lambda^3F^*$
and $\Lambda^6F^*$ parts of $\rep{133}$ on $X$. Identifying
\begin{equation}
\label{eq:AA'def}
\begin{aligned}
   \mu_{mnp8} &= \tfrac{1}{2}A_{mnp} \in \Lambda^3V^* \\
   \mu^m{}_8 &= - A^{\prime m}_{1\dots 7} \in \Lambda^6V^*
\end{aligned}
\end{equation}
where $A^{\prime m}_{p_1\dots p_7}
=(7!/6!)\delta^m_{[p_1}\tA_{p_2\dots p_7]}$ we have the action in the
Lie algebra
\begin{equation}
\label{eq:AA'action}
   (A+\tA)\cdot X = i_xA
      + \big(i_x\tA + A\wedge\omega\big)
      + \big(jA\wedge\sigma - j\tA\wedge\omega\big) .
\end{equation}
Here we have introduced a new notation. The symbol $j$ denotes the
first pure $F^*$ index of sections of $F^*\otimes
(\Lambda^7F^*)$. Hence
\begin{equation}
\label{eq:jdef}
   \left(j\alpha^{(p+1)}\wedge\beta^{(7-p)}\right)_{m,n_1\dots n_7}
      := \frac{7!}{p!(7-p)!}
         \alpha_{m[n_1\dots n_p}\beta_{n_{p+1}\dots n_7]}
\end{equation}
%


\subsection{The $\hg$ subgroup and spinor indices}
\label{app:SU8}

The maximal compact subgroup of $\E7$ is $\hg$. In the
supersymmetry transformations, the spinors transform in the
fundamental representation under (the double cover) $\SU(8)$. Thus it
is often useful to have the decomposition of the various $\E7$
representations in terms of $\hg$.

In particular, one can use the common $\Spin(8)$ subgroup to
relate the decompositions under $\SL(8,\bbR)$ and $\hg$. Let
$\tgam^{ab}$ be $\Spin(8)$ gamma matrices defined
in~\eqref{eq:gamma8def}. We can raise and lower $\SL(8,\bbR)$ indices
using the metric $\tg$. Similarly, spinor indices can be raised a
lowered using $C^{-1}$. Under $\hg$ the $\rep{56}$ representation
decomposes as
\begin{equation}
\begin{aligned}
   X &= ( x^{\alpha\beta}, \bar{x}_{\alpha\beta} ) \\
   \rep{56} &= \rep{28} + \rep{\bar{28}} .
\end{aligned}
\end{equation}
If the symplectic product takes the form
\begin{equation}
\label{eq:OdefSU8}
   \Omega(X,Y) = \ii \left( x^{\alpha\beta}\bar{y}_{\alpha\beta}
      - \bar{x}_{\alpha\beta}y^{\alpha\beta} \right)
\end{equation}
then $(x^{ab},x'_{ab})$ and $(x^{\alpha\beta},\bar{x}_{\alpha\beta})$
are related by
\begin{equation}
\begin{aligned}
   x^{\alpha\beta} &= \tfrac{1}{4\sqrt{2}}(x^{ab}+\ii x^{\prime ab})
      \tgam_{ab}{}^{\alpha\beta} , \\
   \bar{x}_{\alpha\beta} &= \tfrac{1}{4\sqrt{2}}(x^{ab}-\ii x^{\prime ab})
      \tgam_{ab\,\alpha\beta} ,
\end{aligned}
\end{equation}
with $\tgam_{ab}{}^{\alpha\beta}$ and $\tgam_{ab\,\alpha\beta}$ given
by~\eqref{eq:bigammas}, or equivalently
\begin{equation}
\label{eq:SLSU}
   \begin{pmatrix}
      x^{\alpha\beta} \\ \bar{x}_{\alpha\beta}
   \end{pmatrix} = \frac{1}{4\sqrt{2}}
       \begin{pmatrix}
          \tgam_{ab}{}^{\alpha\beta} & \ii\tgam^{ab\,\alpha\beta} \\
          \tgam_{ab\,\alpha\beta} & - \ii\tgam^{ab}{}_{\alpha\beta}
       \end{pmatrix}
       \begin{pmatrix}
          x^{ab} \\ x'_{ab}
       \end{pmatrix}
\end{equation}
Recall that the $\SU(8)$ subgroup of $\Cliff(7,0;\bbR)$ leaves the
norm $\bar{\ksp}{\ksp}$ invariant. Since the defining $\rep{56}$
representation decomposes as a spinor bilinear, both the $\id$ and
$-\id$ elements in $\SU(8)$ leave $X$ invariant and hence we see
explicitly that the subgroup of interest of $\E7$ is actually $\hg$.

Viewing a 56-dimensional index either as a pair of $\SL(8,\bbR)$
indices or as a pair of spinor indices, the relation~\eqref{eq:SLSU}
can be used to convert between $\SL(8,\bbR)$ and $\hg$ decompositions
of any other $\E7$ representations. In particular, decomposing the
adjoint representation $\mu$ under $\hg$ as
$\rep{133}=\rep{63}+\rep{35}+\rep{\bar{35}}$ and writing
$\mu=(\mu^\alpha{}_\beta,\mu^{\alpha\beta\gamma\delta},
\bar{\mu}_{\alpha\beta\gamma\delta})$, with
$\bar{\mu}_{\alpha\beta\gamma\delta}=\smu_{\alpha\beta\gamma\delta}$
and $\mu^\alpha{}_\alpha=0$, one finds
\begin{equation}
\begin{aligned}
   \delta x^{\alpha\beta} &=
     \mu^\alpha{}_\gamma x^{\gamma\beta} +
     \mu^\beta{}_\gamma x^{\alpha\gamma} +
     \mu^{\alpha\beta\gamma\delta} \bar{x}_{\gamma\delta} , \\
   \delta\bar{x}_{\alpha\beta} &=
     - \mu^\gamma{}_\alpha \bar{x}_{\gamma\beta} -
     \mu^\gamma{}_\beta \bar{x}_{\alpha\gamma} +
     \bar{\mu}_{\alpha\beta\gamma\delta} x^{\gamma\delta} ,
\end{aligned}
\end{equation}
with
\begin{equation}
\label{eq:SU8decomp}
\begin{aligned}
   \mu^\alpha{}_\beta
       &= \tfrac{1}{4}\mu^-_{ab}\tgam^{ab\,\alpha}{}_\beta
         - \tfrac{1}{48}\ii\mu^-_{abcd}\tgam^{abcd\,\alpha}{}_\beta , \\
   \mu^{\alpha\beta\gamma\delta}
       &= \tfrac{1}{16}\big( 2\mu^+_{ac}g^{(8)}_{bd}
          + \ii\mu^+_{abcd} \big)
          \tgam^{ab\,[\alpha\beta}\tgam^{cd\,\gamma\delta]} .
\end{aligned}
\end{equation}
where $\mu^\pm_{ab}=\tfrac{1}{2}(\mu_{ab}\pm\mu_{ba})$ and
$\mu^\pm_{abcd}=\tfrac{1}{2}(\mu_{abcd}\pm *\mu_{abcd})$, and the
anti-symmetrisation in the second line is only over $\alpha$, $\beta$,
$\gamma$ and $\delta$. Note that in both lines the contraction with
the relevant combination of gamma matrices automatically projects onto
$\mu_{ab}^\pm$ and $\mu_{abcd}^\pm$, so one could in practice leave
out the $\pm$~superscripts.

Finally we can similarly introduce the decomposition of the
$\rep{912}$ representation under $\hg$, writing
$\phi=(\phi^{\alpha\beta},\phi^{\alpha\beta\gamma}{}_\delta,
\bar{\phi}_{\alpha\beta},\bar{\phi}_{\alpha\beta\gamma}{}^\delta)$. The
adjoint action of $\E7$ then takes exactly the same form as
in~\eqref{eq:muphi} but with spinor indices replacing $\SL(8)$
indices. We can also write the $\rep{56}\times\rep{912}\to\rep{133}$
product in terms of spinor indices. One find
\begin{equation}
\label{eq:XphiSU8}
\begin{aligned}
   \mu^\alpha{}_\beta &= \tfrac{3}{4}\ii\left(
         x^{\alpha\gamma}\bar{\phi}_{\gamma\beta}
         - \bar{x}_{\beta\gamma}\phi^{\gamma\alpha}\right)
      + \tfrac{3}{4}\ii\left(
         x^{\gamma\delta}\bar{\phi}_{\gamma\delta\beta}{}^\alpha
         - \bar{x}_{\gamma\delta}\phi^{\gamma\delta\alpha}{}_\beta
         \right) , \\
   \mu_{\alpha\beta\gamma\delta} &= - 3\ii \left(
         \bar{\phi}_{[\alpha\beta\gamma}{}^\epsilon
            \bar{x}_{\delta]\epsilon}
         + \tfrac{1}{4!}
            \epsilon_{\alpha\beta\gamma\delta \mu_1\dots \mu_4}
            \phi^{\mu_1\mu_2\mu_3}{}_\epsilon x^{\mu_4\epsilon} \right) .
\end{aligned}
\end{equation}
The additional factor of $\ii$ as compared to the $\SL(8)$
expression~\eqref{eq:Xphi} comes from the addition factor of $\ii$
between the $\SL(8)$ and $\hg$ expressions for the symplectic
form~\eqref{eq:Odef} and~\eqref{eq:OdefSU8} respectively.



\end{document}